\documentclass[twocolumn,prl]{revtex4}

\usepackage{amsmath}
\usepackage{amssymb}
\usepackage{latexsym}

\newenvironment{extract}{%
\par\leftskip1pc\rightskip1pc\small}{\par\leftskip0pc\rightskip0pc}

\begin{document}

\title{Life is physics: evolution as a collective phenomenon far from equilibrium}

\author{Nigel Goldenfeld$^1$ and Carl Woese$^{1,2}$}

\affiliation{$^1$Department of Physics, Center for the Physics of
Living Cells, and Institute for Genomic Biology, University of Illinois
at Urbana-Champaign, 1110 West Green St., Urbana, IL 61801,
USA\\$^2$Department of Microbiology and Institute for Genomic Biology,
601 South Goodwin Avenue, Urbana, IL 61801, USA }

\begin{abstract}
Evolution is the fundamental physical process that gives rise to
biological phenomena.  Yet it is widely treated as a subset of
population genetics, and thus its scope is artificially limited.  As a
result, the key issues of how rapidly evolution occurs, and its
coupling to ecology have not been satisfactorily addressed and
formulated.  The lack of widespread appreciation for, and understanding
of, the evolutionary process has arguably retarded the development of
biology as a science, with disastrous consequences for its applications
to medicine, ecology and the global environment. This review focuses on
evolution as a problem in non-equilibrium statistical mechanics, where
the key dynamical modes are collective, as evidenced by the plethora of
mobile genetic elements whose role in shaping evolution has been
revealed by modern genomic surveys.  We discuss how condensed matter
physics concepts might provide a useful perspective in evolutionary
biology, the conceptual failings of the modern evolutionary synthesis,
the open-ended growth of complexity, and the quintessentially
self-referential nature of evolutionary dynamics.

\end{abstract}

\maketitle


\section{Introduction}

Shortly after the publication of the draft of the Human
Genome\cite{lander2001initial}, Stephen Jay Gould, the noted
evolutionary biologist, penned an extraordinary Op-Ed piece in the New
York Times. Although the occasion was in some sense the apotheosis of
the molecular biology revolution, begun nearly 50 years earlier with
the elucidation of the structure of DNA by Watson and Crick at the
Cavendish Laboratory, Gould's tone was far from laudatory.  For, with
the entire genome in hand, the genes could be immediately counted,
yielding an initial estimate of between 30,000 and 40,000 genes, half
as many again as those in the genome of the tiny roundworm {\it C.
elegans} (we now know that the initial estimate was incorrect, and the
correct number of genes is smaller, closer to 21,000\cite{CLAM07}).
Gould commented:

\begin{extract}
\lq\lq Homo sapiens possesses between 30,000 and 40,000 genes... In other
words, our bodies develop under the directing influence of only half
again as many genes as the tiny roundworm ....The collapse of the
doctrine of one gene for one protein, and one direction of causal flow
from basic codes to elaborate totality, marks the failure of
reductionism for the complex system that we call biology ... The
key to complexity is not more genes, but more combinations and
interactions generated by fewer units of code — and many of these
interactions (as emergent properties, to use the technical jargon) must
be explained at the level of their appearance, for they cannot be
predicted from the separate underlying parts alone."\cite{GOUL01}
\end{extract}

In other words, if we are to understand our biology, we need a
statistical mechanics of genes: the fundamental processes that have
shaped us are strongly collective in nature, and need to be treated
appropriately.

The complexities of the human genome are by no means an isolated
example of collective phenomena in biology.  The majority of cellular
life is microbial, and these organisms also are strongly interacting.
Microbes are able to exchange genes (horizontal gene
transfer)\cite{ochman2000lgt}, communicate between cells (quorum
sensing)\cite{BASS05}, translocate collectively over surfaces (swarming
motility)\cite{HARS03,COPE09}, and form biofilms---spatially-extended
multicellular colonies with coordinated division of labor, cellular
differentiation and cooperative defense against
antagonists\cite{SHAP98}.

Clearly, collective processes abound in biology, but they have been
relatively neglected by the biological physics community; most of the
biological research being performed in physics departments today is an
outgrowth of the tremendous technological advances in molecular
biology, single-molecule biophysics and computational biology.

Also neglected, but with even less justification, is the process
of evolution itself; written off as a solved problem under the
catch-phrase \lq\lq natural selection", it was relegated to a
peripheral role during the development of molecular
biology\cite{WOES09}.  With the growing recognition of the importance
of collective phenomena in evolution
especially\cite{herz1994collective,nowak2006evolutionary,GOLD07,shapiro2010mobile},
but also in
ecology\cite{levin1992problem,toner2005hydrodynamics,sumpter2006principles,
schweitzer2007brownian,HALO08},
immunology\cite{cohen2007explaining,cohen2007real},
microbiology\cite{ben2000cooperative,WING06,velicer2009evolution} and
even global climate change\cite{LENT08,BARD08,HAST09}, it is timely to
assess the extent to which a condensed matter physics
perspective---with its unifying principles of collective behavior
arising from interactions---can be illuminating in biology.  Equally
fascinating is the notion that biology may extend the frontier of
non-equilibrium physics, revealing principles of self-organization that
seem absent in purely physical processes such as pattern formation. Is
the study of biology merely an exercise in reverse-engineering a rich
set of extremely complicated chemical reactions, or are deeper
principles at work to bring molecules to life?

The purpose of this article is to discuss these questions by providing
a necessarily selective survey of evolutionary biology, highlighting
the role of collective effects and thereby the possible connections
with condensed matter physics.  Thus, no attempt is made here at a
complete review of evolution.  Instead, we have tried to focus on
points of principle, where the shortcomings of present-day
understanding seem most acute to us.  It would be impossible to review
the entire literature of evolution or even the sub-set of this
literature that is potentially engaging for physicists. For a useful
and self-contained introduction to the conventional, eukaryote-centric
framework of evolutionary theory, accessible to physicists with little
biological background, we refer the reader to the standard text by
Maynard Smith\cite{SMIT93}; an introduction that also covers
physics-related topics especially has been given by
Drossel\cite{DROS01}.  Most of life is microbial, and a modern
microbe-centric view of evolution can be found in the book by
Sapp\cite{SAPP09}.

We envisage a readership with a wide range of knowledge and interest in
evolutionary biology. To the biologist interested in practical issues,
we ask that you do not dismiss the seemingly useless and na\"{\i}ve
issues that we necessarily raise. On the one hand, a fundamental understanding of
evolution may not seem to offer immediate benefits in terms of finding
the next wonder drug; on the other hand, the lack of appreciation for
the rapidity and pervasiveness of evolution has, within a lifetime,
destroyed the effectiveness of numerous
antibiotics\cite{salyers1997arg}, and probably is responsible for the
limited success of the treatment of cancer\cite{gatenby2009change}. The
biomedical-industrial complex cannot afford to ignore the need to
create a fundamental science of biology.  To the physicist, who might
be repelled by the seeming lack of structure represented by biology, we
ask that you look beyond the currently incomplete state of biological
understanding, and be open to the possibility that evolution is both a
physical phenomenon and the natural framework in which biology is
embedded.  The lack of structure in the way that biology is
traditionally presented reflects the field's unavoidable focus on a
single sample path; however, the underlying evolutionary process itself
is surely one with deep mathematical structure, capable of expression
and elucidation in unifying terms based on emergent physical laws. This
is a true frontier of physics, but one that will require a great deal
of what has been termed (in another context of non-equilibrium physics)
\lq\lq open-minded spadework"\cite{LANG80} to unearth.

\section{Evolution as a problem in condensed matter physics}

\subsection{Life is Chemistry}

Most living organisms are composed of cells.  Setting aside, for the
moment, the question of what constitutes \lq\lq life", one might
attempt to gain further insight into the nature of life by examining
carefully the contents of cells. This enterprise culminated in the
advent of molecular biology in the second half of the twentieth
century, and it is now firmly established that cells contain a
multitude of molecular components to perform the functions of life.
These include such examples as: phospholipid bilayers to bound the
cell; cytosol containing water and organic molecules that form the
fluid matrix of the cell; nucleic acids to record genetic information;
proteins comprised of  strings of amino acids that inter alia structure
the cell, pump molecules in and out of the cell, catalyze metabolic
processes, perform mechanical functions, participate in the cell cycle,
transcribe and translate genetic information, and allow for signaling
between cells; and small molecules such as adenosine triphosphate that
transport energy within a cell. The study of life must proceed by the
study of the machinery within the cell, with additional consideration
given to multi-cellular phenomena where applicable.  Moreover, any
discussion of the origin of life from early geochemistry should focus
on the chemical environments from which life might have arisen, and
describe the biochemical pathways through which metabolic and genetic
information could arise spontaneously from such a milieu.  Indeed, as
van Helmont concluded in 1648, and as is even today the rallying cry at
conferences on the origin and evolution of life, it seems quite clear
that \lq\lq  all life is chemistry"\cite{VANH48}.

Or is it?

\subsection{Life is Physics}

From a condensed matter physicist's perspective, we have
been led to a rather strange conclusion.  To see why this is the case,
and why it is important to dwell on this point, let us reflect on the
modern way in which many readers of {\it Annual Reviews of Condensed
Matter Physics\/} view their field.  We will contrast that with an
explicitly reductionist perspective, one which typifies most (but
certainly not all) reasoning in biology.  To begin, suppose that we
were to apply this \lq\lq biological" perspective to study the
phenomenon of superconductivity.  The first step might be the
construction of a catalogue of the known superconductors, a list that
might include elements such as Niobium or Tin, the cuprate oxides such
as La$_{2-x}$Sr$_x$CuO$_4$, the alkali-doped fullerenes such as
Cs$_2$RbC$_{60}$, and the heavy-fermion materials such as
URu$_2$Si$_2$.  From consideration of the structure of the electronic
band structure of these materials, the biologist might try to argue,
quite reasonably, that the transfer of electrons between outer atomic
orbitals is somehow the cause of the interesting transport,
thermodynamic and electrodynamic response of these materials.  Even
though we would not be able to actually construct a predictive theory
of superconductivity, we might still conclude that \lq\lq  all
superconductivity is chemistry". Although we have referred to this
hypothetical line of reasoning as a \lq\lq  biological perspective", it
was actually the approach tried originally by physicists, an approach
that failed.  In fact, such a line of reasoning was attempted by A.
Einstein in 1922\cite{EINS22} (see also \cite{sauer2007einstein} for
historical context) prior to the development of quantum theory in its
mature form. From the manifest failure of his effort, Einstein
concluded (correctly) that not only was superconductivity of quantum
mechanical origin, but that it must involve electrons transported
between closed chains of orbitals, an uncanny precursor of Feynman's
theory for macroscopic quantum order in superfluid
helium\cite{feynman1953atomic}. We learn from all this that
reductionism is a natural, intuitive step in the construction of
theory, and that its failure mode can be an instructive pointer to the
necessary ingredients of a successful theory.

The fallacy of \lq\lq  superconductivity is chemistry" is more evident
than the fallacy in the logic leading to the conclusion that \lq\lq
life is chemistry".  We know now, starting with the work of Ginzburg
and Landau, that superconductivity has, in fact, little to do with the
quantum chemistry of the atoms in the material.  Instead,
superconductivity is best understood as arising from the breaking of
the global U(1) gauge symmetry in the effective field theory that
describes the interaction between off-diagonal long-range order and the
electromagnetic field.  Any microscopic Hamiltonian whose effective
field theory is Ginzburg-Landau theory will give rise to the phenomena
associated with superconductivity.  There is nothing fundamental about
the atoms or molecules.  Indeed, our putative biologist would be
astonished to learn of color superconductivity in deconfined quark
matter\cite{ALFO08}, likely to be realized in the neutron star remnants
of core-collapse supernovae.  Such high temperature superconductivity
would be inconceivable from a narrow perspective based purely upon
atomic chemistry; moreover, the conceptual relationship between the
astrophysical and condensed matter versions of superconductivity would
appear obscure.

To summarize, the phenomenon of superconductivity as a process is
captured by the universal, symmetry-based Ginzburg-Landau theory, but
that process can have many different realizations or instantiations
from matter at a variety of energy and length scales.  This level of
understanding is one that crucially informs the modus operandi of
condensed matter physics.  Typically, we regard a phenomenon as
essentially understood when two conditions have been met.  The first is
that the reasons for the very existence of the phenomenon are known,
usually from some form of symmetry or topological consideration.  The
second is that we have a way to determine under what circumstances a
particular system (atomic, molecular, nuclear, etc.) represents an
instantiation or realization of the phenomenon.  Thus, the mantra of
condensed matter physics, sometimes attributed to Murray Gell-Mann but
surely anticipated in spirit by Lev Landau, becomes a recipe for
discovery: \lq\lq That which is not forbidden is mandatory".  By this
is meant that any allowable process can occur, and therefore it
behooves one to try and identify the universality classes (i.e.
categories) of interesting phenomena, and then to try and identify the
likely realizations of them.  This pattern of discovery is a relatively
recent one, arising during the emergence and maturation of condensed
matter physics as a scientific discipline during the last 25 years of
the twentieth century, and exemplified, for example, by the new and
important condensed matter sub-field of topological insulators, where
the theoretical understanding led to a concerted and successful search
for experimental realizations.  This example is by no means an isolated
one: other examples of major significance include the Aharonov-Bohm
effect, the quantum spin hall effect, localization, and the renaissance
in atomic, molecular and optical physics provided by the experimental
realization of atomic Bose-Einstein condensates.

\subsection{The Need for a Physics of Living Systems}

The point of this lengthy discourse should now be clear.  It is as
limiting to view life as chemistry as it is to view condensed matter
collective states as being primarily about the atoms.  The physics of
living systems would have at its core a generalized description of
evolutionary processes, reflecting allowable dynamical symmetries, not
merely static configurational ones.  Such a description would, in a
very crude sense, be a counterpart to the description of emergent
states of matter embodied by effective theories such as the
Ginzburg-Landau theory alluded to above.  A genuine physics of living
systems would encompass different limits of the evolutionary process,
each recognized as, and described by different effective theories.

A biological example of a universality class and the corresponding
effective theory is classical population genetics, itself sub-divided
into two universality classes, one for sexual organisms and one for
asexual organisms.  Taken at face value, population genetics contains
phenomenological parameters, such as effective population size, fitness
and growth rate, and attempts to model generic aspects of populations
and their genes.  It certainly is not a microscopic theory, because it
lacks a biochemical level of description, and it is certainly an
effective theory, valid only when there is a separation of scales
between ecosystem dynamics and gene mutation dynamics.  Condensed
matter physics was liberated from the hegemony of Fermi liquid theory
by the recognition that there are many other universality classes
describing the behavior of electrons in solids, representing different
types of collective behavior and interactions between (e.g.) spin and
charge degrees of freedom.  Similarly, we anticipate that evolutionary
biology can be liberated from the hegemony of classical population
genetics through the recognition that other universality classes must
exist, and will be manifest under the appropriate conditions. We will
say more below about the ways we see that the discoveries of modern
biology are positioning it to take the next step beyond the paradigm of
classical population genetics.

This perspective redefines what we demand of biological understanding
in two additional ways.  First, the very existence of the phenomenon of
life needs to be understood. Second, the realization or instantiation
of it, on Earth, for example, needs to be understood.  For the most
part, it is fair to say that the discipline of biology has neglected
the first condition, and in pursuit of the second, has confused
understanding of the realization with understanding of the phenomenon.
This has had a number of unfortunate consequences, which arguably have
hindered both the conceptual development of biology and the proper
application of foundational understanding to societal applications.

Let's begin with the conceptual difficulties.  A unified view of a
phenomenon, such as we have alluded to in superconductivity, has the
benefit that further instances of it do not come as a surprise, and do
not require further ad hoc explanation.  With a proper understanding of
superconductivity as a symmetry-breaking process, for example, one does
not find it surprising to learn of superconductivity in nuclear or
astrophysical contexts.  With a proper understanding of the phenomenon
of life as a dynamical process, for example, one would not find it
surprising to learn of life in so-called extreme environments (such as
deep beneath the ocean floor) or even on other planets.  Amongst the
surprises that biology has encountered recently are several of major
significance, including: the discovery of horizontal gene transfer in
multicellular
eukaryotes\cite{gladyshev2008massive,keeling2008horizontal,
palenik2009coastal,monier2009horizontal,pace2008repeated} and the
discovery of a fully anoxic multicellular life form\cite{DANO10}.  In
short, a unified view prevents the unnecessary multiplication of
hypotheses that is the sure sign of a lack of fundamental understanding
(think epicycles!).

The second consequence of a lack of fundamental understanding in
biology is the failure to recognize that biology is a manifestation of
evolution --- not the other way round.  Interventions in biological
systems inevitably provoke an evolutionary response which is rapidly
emerging and spatially-distributed.  Examples of the ability of
biological systems to defeat human attempts at mitigation include: (1)
the world-wide spread of antibiotic resistance genes across distantly-related
bacteria, crossing species and phylum boundaries and physical
locations\cite{salyers1997arg}; (2) the rapid evolution of cancer
tumors in the face of chemical
attack\cite{gatenby2009change,kimmel2010evolution,attolini2009evolutionary};
(3) the ability of HIV to out-adapt treatment\cite{DUFF08,NEHE10}; (4)
the ability of life to adapt to the massive poisoning of the
Precambrian atmosphere by cyanobacteria-released oxygen 2.4 billion
years ago (changing from a reducing to an oxidizing
atmosphere\cite{FALK08}) with a remarkable subsequent flowering of
life.  With the exception of example (4), these examples indicate a
fundamental limitation to medical science, akin to trying to design
integrated circuits without a fundamental knowledge of quantum
electronics and semiconductor physics.

Perhaps the primary shortcoming of the biological enterprise is the
manifest failure to account for the phenomenon of the existence of
life.  Without doubt, this failure reflects not only on biologists, but
also on physicists.  We say this because the majority of biologists
would probably regard their primary role as being in one sense or
another to reverse engineer the myriad specific realizations of organic
life on Earth --- the reductionist exercise that has been notably
successful within its own terms of reference.  On the other hand, the
existence of the phenomenon of life, if it can be understood at all in
generic terms, is surely an emergent phenomenon, arising somehow as an
inevitable consequence of the laws of non-equilibrium statistical
physics.  How is it that matter self-organizes into hierarchies that
are capable of generating feedback loops which connect multiple levels
of organization and are evolvable? When life emerged from early
geochemistry the process must have been driven by irreversible
thermodynamics, but the extension of that process into the emergence of
evolvable structures remains mysterious to us.  The physical laws that
govern far from equilibrium dynamics are still not known.

\subsection{Are there new physical laws in Biology?}

In 1949, Delbr\"uck, inspired by Bohr, famously expressed the view that
biology might exhibit phenomena that are beyond quantum mechanics:

\begin{extract}
\lq\lq Just as we find features of the atom, its stability, for instance,
which are not reducible to mechanics, we may find features of the
living cell which are not reducible to atomic physics, but whose
appearance stands in a complementary relationship to those of atomic
physics." --- M. Delbr\"uck, A Physicist Looks at Biology (1949).
\cite{DELB49}
\end{extract}

\medskip
Today, few seriously expect that such features or physical laws will be
found.  However, in the same essay, in a less-celebrated passage,
Delbr\"uck drew attention to the problem of spontaneous generation of
life.  Surprisingly, to him the interesting issue was how statistical
fluctuations in the kinetics of the emergence of life would lead to a
possible lack of determinacy in the biochemistry of living organisms:
the organism might not precisely reflect the geochemistry from which it
arose.  Delbr\"uck was writing at a time when phase transitions were only
beginning to be understood, and the notions of emergence and
spontaneous symmetry breaking were in their infancy.  Thus, he probably
had no precise notion of a \lq\lq law" of physics being a description
of an effective theory, one that is valid on an intermediate asymptotic
scale of length, time or energy, and systematically related to a deeper
level of description.  Whereas Delbr\"uck looked to biology to extend
quantum mechanics, we look to it as a source of insight into
non-equilibrium statistical mechanics of the evolutionary process.

Thus, the study of biology should be more than simply cataloguing the
wonders of biological organization.  We see no reason why the mantra
attributed to Gell-Mann should not apply with equal force and
predictive power in biology, and become part of its methodology.  Today
a condensed matter physicist envisages a new class of collective
processes and finds realizations in the world of materials, or for that
matter, optical lattices.  What would biology look like as a science if
we sought to anticipate the types of evolutionary processes available
to the suite of genetic operators now known?  What realizations of such
processes could we find, if we simply looked?

Maybe this sounds far-fetched, but in fact this methodology has already
been used to good effect in biology.  One example we have in mind is
the groundbreaking discovery of the class of molecules known as
topoisomerases, whose existence was first recognized theoretically,
leading to their eventual discovery\cite{wang98,WANG09}.  The
topoisomerases are a class of enzymes that are able to perform the
miracle of passing one strand of DNA through another, by breaking and
then reforming them. The end result of this process is that DNA can be
uncoiled through a process of successive topological changes, allowing
transcription and replication to occur.  Without the topoisomerases,
this would be essentially impossible on the timescale relevant to
cellular processes.

\section{Beyond the Modern Synthesis}

The classical and widely-accepted framework of evolution is the
so-called \lq\lq  Modern Synthesis" or \lq\lq Neo-Darwinism", based on
the fusion of Mendelian genetics with Wallace\cite{WALL1858} and
Darwin's ideas\cite{DARW1859} about \lq\lq  natural selection" (or
\lq\lq survival of the fittest", the terminology preferred by
Wallace)\cite{HUXL42,SMIT93}. This theory and additions to it,
primarily those due to Kimura\cite{KIMU85}, account for very simple
genetic processes, such as point mutation and sexual recombination,
leading to random single nucleotide polymorphisms.  A characteristic of
these classical theories of evolution is that their genome dynamics is
linear, diffusive in nature, and the population sizes of communities
are typically sufficiently large that fixation times are long.  This
union of evolution and genetics that developed in the 1930's and 1940's
presumed that evolution proceeded through the simple mechanism of
heritable mutation and survival of the fittest. Organisms have
offspring that survive and propagate based on the quality of the
inevitably mutated genome they inherited from their parents (vertical
gene transfer).  New positive traits spread through the population
because the individuals with those traits were more successful in
surviving and breeding than other members of the population.  A variety
of possible niches in the physical environment enable the
multiplication of species, which then interact to create new niches.
Thus, in this picture, evolution is essentially synonymous with
population genetics.  Genes are assumed to be the only dynamical
variables that are tracked, and are associated with a fitness benefit
that is difficult to define or measure precisely, but is quantified by
a fitness landscape that describes how the population fitness depends
on the genotype\cite{wright1932rmi, gavrilets2004fla, orr2005gta,
PARK10}. Traits are simply associated with genes, and gene interactions
are often ignored, or at best handled through the fitness
landscape\cite{NEHE09,PARK10}.

Implicit in this approach is the assumption that the evolutionary
time-scale is different from the time-scale of the ecosystem. The
crucial question of the timescale of the evolutionary process, even
taking at face value the perspective of the modern synthesis (which we
do not) remains a thorny issue\cite{GING09}, and indeed it is fair to
say that the theory's conceptual framework is so poorly quantified that
one cannot confidently make sensible and realistic estimates of
timescales (for an excellent pedagogical discussion of this point, see
\cite{FISH07}).

\subsection{Epistasis}

Even within the framework of the Modern Synthesis, one can begin to
make model calculations of how evolution rates vary with population
size and the nature of the fitness
landscape\cite{desai2007speed,brunet2008stochastic,
fogle2008clonal,weissman2009rate,PARK10}, and probe the role of collective
effects---epistasis---between genes.  For example, phenotypic variance
is generally thought to be the result of many-gene
interactions\cite{MOOR05}, as documented in a tour de force analysis of
the yeast metabolic network, for example\cite{SEGR05}.  In yeast, a
single-cell eukaryote, any positive selection that accrues from many
interacting genes with small but positive contributions to fitness must
contend with recombination that mixes up and randomizes the genotype.
This competition can lead to collective states, more or less by analogy
to phase transitions: a high recombination rate (analogous to high
temperature) where genes are weakly correlated and genotypes are
short-lived, and a low recombination rate (analogous to low
temperature) regime where favorable genotypes are stable and compete
essentially clonally\cite{NEHE09}.  In bacteria, which reproduce
clonally, recombination is a stabilizing factor and can compete in a
similar way with point mutations, leading to two phases: one that
contains a narrow distribution of genotypes and one that is genomically
diverse, with global genome sequence divergence arising through the
propagation along the genome in evolutionary time of diversification
fronts triggered by horizontal gene transfer events\cite{VETS05}, or
perhaps even by indels\cite{CHEN09}.  How these dynamics play out when
the spatial structure of microbial populations is included is a
fascinating question, bringing together genome dynamics, ecosystem
dynamics and population dynamics in a way that has not yet been
explored.

\subsection{Mobile genetic elements}


The difficulties in making detailed and quantitative theories of the
rate of evolution become vastly more acute as a result of discoveries
from the emerging science of genomics\cite{shapiro2010mobile}. Building
on the seminal work of Barbara McClintock\cite{MCCL84}, the past decade
or so has witnessed the discovery of a plethora of what one might term
\lq\lq classically-forbidden" processes that radically transform our
understanding of dynamics at the level of the genome\cite{SHAP09}. Of
particular importance is the discovery of mobile genetic elements in
many forms, ranging from transposons to horizontal gene transfer
agents\cite{syvanen1994hgt,ochman2000lgt,filee2003rpv,weinbauer2004vdm,frost2005mobile,babic2008direct,PAL07},
whose levels of activity and evolutionary impact have almost surely been
severely underestimated\cite{PAUL10}. Horizontal transfer means that
genes or other genetic materials are transmitted through a variety of
non-hereditary mechanisms from one organism to another, and
subsequently expressed, thus altering the behavior (phenotype) of the
recipient.  Long known to be present in Bacteria and Archaea,
horizontal gene transfer is now known to be present in multicellular
Eukaryotes as well, as a result of genome-wide surveys published in the
last year or
so\cite{HOTOPP07,gladyshev2008massive,keeling2008horizontal,palenik2009coastal,monier2009horizontal,pace2008repeated}.
While the horizontal transfer of genes is widely recognized to be a
major evolutionary force in Archaea and Bacteria, it is still too early
to be precise about its role in Eukaryotes.

However, it is not only genes that can be transferred: in Eukaryotes,
genes are a small fraction of the total genome, with non-coding DNA and
transposable elements making up the majority of the genome in some
cases.  Transposable elements, sometimes inaccurately but colorfully
known as \lq\lq jumping genes", can easily move around and between
chromosomes, and through the disruption that they potentially inflict upon
a genome, cause deleterious mutations and illness.  In humans, it is
estimated that about 45\% of the genome is composed of transposable
elements\cite{lander2001initial}.  Horizontal transfer of transposable
elements can be a major driver of eukaryotic genome evolution and a
source of genetic
innovation\cite{biémont2006genetics,schaack2010promiscuous,rankin2010traits}.
Indeed, the textbook picture of a static genome composed of genes and
junk DNA has now been superseded by recent findings, and it is arguably
more appropriate to think of the genome as a set of one-dimensional
ecosystems, coupled together by horizontal transfer, and containing
numerous genetic elements interacting with each other, creating niches
for themselves, and evolving stochastically to create a community
ecology\cite{venner2009dynamics}.

These genetic mechanisms permit the spread of genetic novelty much
faster than vertical or hereditary transmission of genes, essentially
amounting to a Lamarckian\cite{LAMA1809} dynamic of
evolution\cite{GOLD07,KOON09}.  If this was not enough, it is now
incontrovertible that the inheritance of acquired characteristics, long
discredited, but violating no known law of nature, can sometimes occur,
not only through horizontal gene transfer (e.g. in microbes), but
through so-called epigenetic mechanisms that by-pass the usual modes of
inheritance\cite{JABL98,RAND07} (especially in ciliates\cite{NOWA09}).

Not only is the Modern Synthesis afflicted by strong interactions, but
its very foundation is questionable.  The evident tautology embodied by
\lq\lq survival of the fittest" serves to highlight the
backwards-looking character of the fitness landscape: not only is it
unmeasurable a priori, but it carries with it no means of
expressing the growth of open-ended complexity\cite{GUTT08} and the
generation of genetic novelty.  Thus, the Modern Synthesis is, at best,
a partial representation of population genetics, but this on its own is
a limited subset of the evolutionary process itself, and arguably the
least interesting one.

\subsection{Coupling between evolution and ecology}

It is not only the microscopic basis for current evolutionary theory
that has been challenged by recent advances in biology.  There is now a
substantial and growing literature that documents a surprisingly rapid
rate of evolution in numerous
systems\cite{caporale2003darwin,HAIR05,cowen2005hsp90,
thompson1998rapid,carroll2007evolution,steinhauer1987rapid,zhang2008rapid,summers2008molecular},
ranging from cancer tumors and the immune system to ecosystem-driven
adaptations in all three domains of life.  Moreover, detailed
observations document the coupling between evolutionary and ecological
timescales\cite{HAIR05,meyer2006prey,hillesland2009experimental,
palkovacs2009experimental,bailey2009genes,pelletier2009eco,jones2009rapid,
terhorst2010evolution}.  In a predator-prey system realized in
rotifer-algae interactions, the rapid evolutionary dynamics is
responsible for the unusual phase-lag characteristics of the observed
population oscillations\cite{yoshida2007cryptic}.

Evolutionary and ecological timescales can also become coupled if the
ecological timescale becomes very long: an important example of this is
provided by the ecology of the translation apparatus in the cell.  The
genetic code---the map between triplet codons and amino acids is
degenerate: there can be several synonomous codons that code for the
same amino acid.  It is well-established that the synonomous codons are
not used with equal frequency, and this codon usage bias reflects
selection for speed or accuracy of translation of highly-expressed
genes\cite{BULM87,BULM91}.  The translation process is a highly-complex
one, but relies essentially on the availability of resources, in
particular tRNA molecules.  Evolution of the genome can of course also
lead to a coevolution of the abundance of tRNA in the cell, leading to
a nonlinear dynamics of the genome and its tRNA abundance
distribution\cite{VETS09}.  By going beyond the classical
mutation-drift-selection framework into the regime of nonlinear
evolutionary dynamics, the theory predicts multi-stability of the
genome and an explanation of the pattern of observed microbial genome
biases, not only in translation, but also in transcription and
replication.

The coupling between environmental and ecological timescales has also
been argued to lead to another generic feature of biology: the
prevalence of modularity\cite{HART99,BARA04,KASH05,KASH06}.  Modularity
refers to the relative independence of a biological component or
network---relative, because everything is connected, of course, but the
intramodule connections are more important than intermodule
connections.  Modularity carries with it the connotation of reuse of
motifs, simple building blocks from which complex systems can be
built\cite{HART99}.  One might think that such networks could be
generically obtained from simulations of the evolutionary process, for
example using genetic algorithms and digital life
simulations\cite{HOLL92}, but remarkably this is not the
case\cite{KASH05}.  The reason is that typically such simulations
emulate the assumed process of \lq\lq natural selection": networks are
evolved by mutations, recombination and other genetic operators, and
only those which perform a defined goal well enough are permitted to
enter the next generation. The key to modularity seems to be the
coupling to the environment, as evidenced by two rather different
calculations.  In the first, the network was evolved in an environment
of goals that changed in a modular fashion\cite{KASH05}; moreover, a
follow-up study\cite{KASH06} showed clearly that environmental
fluctuations do indeed accelerate the rate of evolution.  In the second
study\cite{EARL04,HE09}, modularity emerged spontaneously as an outcome
of horizontal gene transfer in the presence of environmental
fluctuations.  This result, admittedly obtained in a rather specific
model, nevertheless highlights the importance of collective
interactions and the interplay between environmental fluctuations and
evolution, neglected in the modern synthesis.

How does horizontal gene transfer influence the architecture of actual
biological networks?  Comparative genomics, coupled with the flux
balance analysis of the metabolic network of {\it E. coli\/} has
demonstrated that the network grows by acquiring genes individually and
in groups (typically operons governing coupled reactions), which are
attached preferentially at the edges of the existing
network\cite{PAL05}. This form of network organization does not
necessarily imply modularity, but it suggests one way that modularity
can arise.  Although there is no unique measure of modularity, the
available analyses indicate that modularity is an increasing function
of the variability of the environment\cite{PART07} and that modularity
also reflects the number of niches available and is associated with
horizontal gene transfer\cite{KREI08}.

An example of the non-trivial coupling between evolution and ecology
has been obtained by recent metagenomic surveys of marine microbial
environments, sampling and analysing environmental DNA collected from
the Sargasso Sea and the Red Sea.  Thirty percent of global carbon
fixation occurs through the photosynthetic pathways of two
cyanobacteria, {\it Prochlorococcus\/} and {\it Synechococcus}.
Remarkably, the phages of these organisms also contain photosystem II
genes, presumably to maintain the host as a functioning
phage factory, thereby increasing the production of phages during the
lysis process as the host cell is destroyed. From analysing the
molecular sequences of these genes, and reconstructing their
evolutionary history, Chisholm's group at MIT have documented the
transfer of photosystem II genes back and forth between these marine
cyanobacteria and their phages\cite{sullivan2006pae}. Moreover, these
genes underwent evolution and sequence shuffling while resident in the
phages.  Thus, rather than supporting the traditional view of the
relationship between microbes and viruses as being a predator-prey
relationship, the new findings suggest that there are collective
interactions between microbes and viruses through gene exchange, with the
creation of an effective global reservoir of genetic diversity that
profoundly influences the dynamics of the major marine ecosystems.
These findings had been anticipated many years earlier by numerous
investigators\cite{ANDE66,ANDE70,sonea1988bwl,syvanen1994hgt,weinbauer2004vdm},
who appreciated and rediscovered the collective outcome of horizontal
gene transfer.

During the last few years, an even more astonishing example has come to
light, prompted in part by the attempt to find the cause of colony
collapse disorder---the dramatic reduction in honey bee population (in
the US, losses of adult workers were 23\% during 2006-7, 36\% during
2007-8)\cite{BERE09}.  One of the potential pathogenic causes, the
Israeli acute paralysis virus (IAPV), was found to be able to integrate
harmlessly its genome into that of the bee host, and thus confer
immunity on the host to further infection.  The surprise is that this
virus is not a retro-virus: it does not need to integrate itself into
the host genome in order to replicate, and so it lacks the genetic
machinery for reverse transcription of its RNA into the host DNA.  It
is not currently known, therefore, how IAPV was able to work its way
into the host genome.  This is not an isolated example: it is now known
that a similar process has occurred in at least 19 vertebrate species,
the relevant viruses that have conferred immunity being the
lethal Bornavirus and Ebolavirus\cite{HORI10,BELY10}.  It seems that
this mechanism is a eukaryotic analogue to lysogeny in microbes. These
findings support the notion that there are collective interactions
between viruses and their hosts.

Evolution and ecology couple not only through time, but also through
space.  Wallace was the first to emphasize that speciation is a
phenomenon localized in both space and time\cite{wallace1855law}:
evolution proceeds through a process of front propagation in space that
couples to population genetics in ways that are conceptually simple but
only now beginning to be understood in a quantitative
way\cite{ibrahim1996spatial,hallatschek2008gene,hallatschek2009fisher,korolev2009}.
As fronts expand, the pioneer organisms at the leading edge experience
large demographic fluctuations that are known to play a significant
role in temporal oscillations\cite{MCKA04} and spatial patterns in
ecosystems\cite{butler2009robust}.  It is important to stress that
horizontal gene transfer is also strongly influenced by spatial
structure.  For example, it was recently established that the frequency
of conjugation events between bacteria is dependent on the local
density, being essentially one/per generation in closely-packed
biofilms, and an order of magnitude smaller in planktonic
culture\cite{babic2008direct}. How the interplay between evolutionary
dynamics, ecosystem dynamics and species distribution is reflected in
patterns of species abundance distributions, diversity measures, and
community structure is a frontier topic in ecology, and relevant to the
emerging conceptual framework of niche construction\cite{ODLI03,DAY03}.

\section{The dynamics of evolution}

Most existing approaches to formulating evolutionary dynamics
mathematically, as Drossel explicitly points out\cite{DROS01}, share
the limitation that the space in which evolution takes place is fixed.
For example, some models consider the dynamics of genomes of fixed
length in a specified fitness landscape, others consider the interplay
between agents who are using specified strategies of behavior in their
repeated encounters with other agents.

Such approaches to evolution miss what is to us the central aspect of
evolution: it is a process that continually expands the space in which
it operates through a dynamic that is essentially self-referential.
Self-reference should be an integral part of a proper understanding of
evolution, but it is rarely considered explicitly.  This point is so
important, because it is at the root of why evolution represents a
non-trivial extension of the sorts of dynamical processes we encounter
in condensed matter physics.  In condensed matter physics, there is a
clear separation between the rules that govern the time evolution of
the system and the state of the system itself.  For example, in
studying fluid dynamics, a firm basis for theory is provided by the
Navier-Stokes equations, and regardless of whether the flow is at low
Reynolds number, dominated by viscous effects, or at high Reynolds
number and dominated by inertial effects, the underlying equations are
capable of capturing all the phenomena.  The mathematical reason for
this is that the governing equation does not depend on the solution of
the equation.  The evolution operator is independent of the state of
the system.  In biology, however, the situation is different.  The
rules that govern the time evolution of the system are encoded in
abstractions, the most obvious of which is the genome itself.  As the
system evolves in time, the genome itself can be altered, and so the
governing rules are themselves changed.  From a computer science
perspective, one might say that the physical world can be thought of as
being modeled by two distinct components: the program and the data. But
in the biological world, the program is the data, and vice versa. For
example, the genome encodes the information which governs the response
of an organism to its physical and biological environment.  At the same
time, this environment actually shapes genomes, through gene transfer
processes and phenotype selection.  Thus, we encounter a situation
where the dynamics must be self-referential: the update rules change
during the time evolution of the system, and the way in which they
change is a function of the state and thus the history of the system.
To a physicist, this sounds strange and mysterious: What is the origin
of this feature that sets biological systems apart from physical ones?
Aren't biological systems ultimately physical ones anyway, so why
is self-reference an exclusive feature of biological systems (whatever
they are!)?

The simple answer seems to be that self-reference arises because the
biological components of interest are emergent, and we are seeking a
description of biological phenomena in terms of these biological
components only. Ultimately, if we used a level of description that was
purely atomistic, for example, this self-referential aspect of biology
would not arise.  This argument does not distinguish biology from
condensed matter physics, where many of the degrees of freedom are
emergent also.

Thus, it is interesting to ask if there are analogues of this
phenomenon in condensed matter physics, generically arising from
coarse-grained description of systems with order parameter dynamics. To
answer this question, recall, for example, the two types of description
that we have of superconductivity.  On one hand there is the BCS theory
of superconductivity, which works at the level of description of
fermions coupled through a pairing interaction (whose microscopic
origin need not be specified, but is due to phonons in classic
superconductors and perhaps spin or other interactions in high
temperature superconductors).  On the other hand, there is the
coarse-grained order parameter description due to Ginzburg and Landau.
Frequently, physicists use the latter as a convenient model that is
easy to calculate with, in systems of arbitrary geometry or with
spatial variation.  Moreover, this level of description is frequently
used to study the time-dependent phenomena in superconductors, given by
the time-dependent Ginzburg-Landau equations. Near the superconducting
critical point, this description is a generic consequence of critical
dynamics.  However, away from the critical region this equation cannot
be systematically derived.  Similarly, in superfluids and Bose-Einstein
condensates, the zero temperature dynamics can be well-described by the
Gross-Pitaevskii equations, and near the critical point, a more
complicated dynamic universality class is believed to be appropriate.
Away from these two regimes, however, there is no universally-agreed
upon description that is systematically accurate. The reason that there
is confusion surrounding the dynamics at intermediate temperatures has
in fact been well-understood, but not widely appreciated, for many
years: the assumption that there exists a description local in both
space and time is false.  This can be seen from the derivation of the
Ginzburg-Landau description from the more microscopic BCS theory, which
in general involves a memory function that becomes local in space and
time only in special limits. The breakdown of locality that accompanies
effective descriptions of dynamical phenomena is well-known beyond
superconductivity, of course, and is a feature of several approaches to
non-equilibrium statistical mechanics, including mode-coupling theory
and renormalization group approaches to effective actions in field
theory.

Why self-reference is a specific feature of biological systems and not
physical systems should now be evident: self-referential dynamics is an
inherent and probably defining feature of evolutionary dynamics and
thus biological systems.  Thus, the question really is how
self-referential dynamics arises as a universality class from the basic
laws of microscopic physics, as an expression of nonequilibrium
physics. Although there is a recognition of this sort of question in
some of the literature on philosophy, artificial life and the evolution
of language, nothing approaching a serious calculation has been done to
our knowledge.

The fact that evolution is a process that transcends its realization
means that evolution is able to act on its own basic mechanisms.  This
nonlinearity of the evolutionary process is sometimes referred to as
evolvability.  It has important generic ramifications that have been
explored in a model calculation of protein evolution\cite{EARL04}, in
particular that evolvability is preferentially selected during periods
of increased rate of environmental change.  Whether or not evolvability
can be selected for, in the conventional parlance, is a controversial
topic, partly because of the unfashionable nonlinearity of the process,
and partly because evolvability seems to undermine the robustness of
biological organization.  Presumably, a detailed quantitative
understanding of evolution would flesh out the balance between
evolvability and robustness\cite{LENS06}.

\subsection{Evolution and Complexity}

Complex systems are characterized by the presence of strong
fluctuations, unpredictable and nonlinear dynamics, multiple scales of
space and time, and frequently some form of emergent structure. The
individual components of complex systems are so tightly coupled that
they cannot usefully be analyzed in isolation, rendering irrelevant
traditional reductionist approaches to science, obscuring causal
relationships, and distinguishing complexity from mere complication.
Biological complexity, is an extreme example of complexity, and arises
from the inclusion of active components, nested feedback loops,
component multifunctionality, and multiple layers of system dynamics,
and is relevant to numerous aspects of the biological, medical and
earth sciences, including the dynamics of ecosystems, societal
interactions, and the functioning of organisms.  Perhaps the most
striking features of biology are the open-ended growth of complexity
that we see in the biosphere, the large population fluctuations, and
the widespread occurrence of the \lq\lq  law of unintended
consequences" when trying to manipulate ecosystems\cite{PALU09}. Thus,
although complexity is hard to define precisely and usefully, we regard
the defining characteristic of complexity as the breakdown of
causality\cite{GOLDeditorial09}. Simply put, complex systems are ones
for which observed effects do not have uniquely definable causes, due
to the huge nature of the phase space and the multiplicity of paths.

Ecosystems are never static but are continually changing and adapting,
and their response involves all levels down to the genome and even
smaller (because viruses are an integral part of an ecosystem).  In an
attempt to capture such complex dynamics, researchers have made
extensive use of digital life simulations.

Digital life
simulations\cite{HOLL92,RAY91,ADAM94,CALD98,DROS01,ADAM02,MCKA04a,MARO06}
use interacting synthetic organisms, with predefined rules for
replication, evolution and interaction with each other and their
digital environment.  Experiments on digital organisms are an accurate
and informative methodology for understanding the process of
evolution because the entire phylogenetic history of a
population can be tracked, something that is much more difficult---but
not impossible\cite{elena2003eem}---to do with natural organisms\cite{ADAM06}.
Experiments on digital organisms can be performed over time scales
relevant for evolution, and can capture universal aspects of
evolutionary processes, including those relevant to long-term
adaptation \cite{wilke2002bdo,LENS99}, ecological
specialization\cite{ADAM95,ostrowski2007esa} and the evolution of
complex traits\cite{LENS03}.  Despite this progress, the way in which
evolution leads to ever increasing complexity of organisms remains
poorly understood and difficult to capture in simulations and models to
date. Is this because these calculations are not sufficiently
realistic, extensive, or detailed, or has something fundamental been
left out?

As early as 1971, Woese speculated on the emergence of genetic
organization\cite{WOES71}.  He was concerned with the evolution of
complexity, but approached from a molecular viewpoint, namely that of
the origin of quaternary structure in proteins.  Woese portrayed
evolution as a cyclic process, in which gene products evolved first to
a dimerized state, followed by gene duplication, and separate evolution
of the gene products so that the dimer finally consists of related but
not identical sub-units.  If the two genes that code for the two
distinct halves of the dimer subsequently fuse, a new composite
molecule has arisen that he called a \lq\lq co-dimer".  In this
co-dimer, the two previously-related components can evolve separately
but in a complementary fashion, as long as the biochemical properties
of the co-dimer as a whole are not adversely affected. In this way, one
half of the co-dimer could, for example, evolve into a control site for
the enzyme that is the other member of the co-dimer.  The point of this
argument is that it provides a molecular realization of the process of
how evolution can cross wide fitness valleys\cite{weissman2009rate}. He
argued that the dynamics of co-dimerization, repeated ad infinitum, led
to the growth of complexity of all macromolecular components of the
cell, including proteins and the translational machinery itself.  If
codimerization were an important part of the evolutionary process, it
would perforce entail there being a high abundance of homodimeric
proteins in the cell, and moreover that the number of subunits would be
an even number. At the time Woese's paper appeared, the paucity of
available data did not permit a test of these predictions.  Early
protein structures were limited to small molecules because of the
difficulty of crystallizing larger molecules, so that when the Protein
Data Bank was surveyed in 1995, 66\% were found to be
monomeric\cite{JONE95}.  By 2000, the situation had dramatically
changed, with 19\% of the proteins surveyed being monomeric. Of the
remainder, 59\% were either dimers or tetramers, with a clear
preference for even numbers of sub-units\cite{GOOD00,PLAX09}.  More
recent analyses confirm these findings and are even able to probe the
evolutionary dynamics that has led to the observed structure of protein
complexes\cite{PERE07}.  Moreover, not only has the detailed structural
evidence consistent with the co-dimerization model been fully
elaborated, but also it appears that the assembly of proteins follows
the evolutionary development of their sub-unit structure\cite{LEVY08}.

Co-dimerization and gene duplication are examples of how biological
systems exploit redundancy as one of the prime mechanisms for
evolution. This insight re-emerges in recent simple models of
evolutionary dynamics, which show that open-ended complexity is only
possible as an outcome of complexity-scale-invariant genomic operators,
such as gene duplication\cite{GUTT08}. That is, if there are genetic
operators that bias organisms to have a specific complexity, then the
complexity of the system will not increase without bound. This
invariance is similar in spirit to that which lies at the heart of the
Richardson cascade in turbulence\cite{richardson1922wpn, KOLM41}.
Guttenberg and Goldenfeld showed in an explicit model of digital
life\cite{GUTT08} how  different genetic operations behaved with regard
to this invariance criterion, and thus were able to devise ecosystem
models that evolved open-ended complexity. Despite its popularity, a
static \lq\lq  fitness landscape" \cite{wright1932rmi,
gavrilets2004fla, orr2005gta} picture of evolution does not satisfy the
proposed invariance criterion, and is indeed conceptually insufficient
to account for the open-ended growth of complexity. Thus, digital life
simulations do not generally evolve qualitatively new responses or
modes of behaviour; they cannot \lq\lq think outside the box".

The emphasis on redundancy as a motif suggests that a component of
evolution is multifunctionalism.  To see why, consider how a
system is modeled, perhaps as a set of differential equations or
lattice update rules. These rules themselves need to evolve: but how?
We need an additional set of rules describing the evolution of the
original rules. But this upper level of rules itself needs to evolve.
And so we end up with an infinite hierarchy, an inevitable reflection
of the fact that the dynamic we are seeking is inherently
self-referential.  The way that the conundrum can be resolved is to
begin with an infinite-dimensional dynamical system which spontaneously
undergoes a sequence of symmetry-breaking or bifurcation events into
lower-dimensional systems. Such transitions can be thought of as
abstraction events: successive lower dimensional systems contain a
representation in them of upper levels of the original system.  A
precise mathematical prototype of such a construction can be
constructed from consideration of the dynamics of functions on a closed
one-dimensional interval\cite{kataoka2003dynamical}, admittedly with
very little direct biological interpretation, but with the positive
outcome of generating a hierarchically entangled dynamical network.
Such networks are not simple tree structures, and this means that the
nodes and links of the network drive each other in a way that a
biologist would interpret as co-evolutionary.  Another way to interpret
such dynamical systems is that the elements are multifunctional: their
input-output map depends on the state of the system, rather than being
a constant in time.  Such systems can have no static fixed point to their
dynamics: in the language of earlier work\cite{GUTT08}, they must
exhibit a complexity cascade, just as Woese had earlier argued in the
context of co-dimerization.

\subsection{Coevolution and game theory}

An alternative to treating organisms as evolving in a fixed environment
or fitness landscape is coevolution\cite{ANDE09}.  In coevolution,
organisms interact and their interactions drive each to evolve, leading
to a continuing process of phenotype evolution, although not
necessarily an increase in complexity, as occurs in the complexity
cascade\cite{GUTT08}.  Examples include mutualism and antagonistic
coevolution.  In mutualism, the interaction is basically symbiotic, and
frequently occurs between plants and animals\cite{BASC07}, microbes and
plants\cite{KIER08} and, gaining increasing attention in the last few
years, humans and their microbiomes\cite{DETH07}).  Antagonistic
coevolution describes how opposing organisms, such as predators and
prey, develop an arms race\cite{van1973new,WEIT05} as a result of their
competition, an effect generally referred to as the Red Queen
effect\cite{van1973new}.

There is a growing empirical
literature\cite{brockhurst2003population,brockhurst2004effect,
fussmann2007eco,gandon2008host,brockhurst2010using,paterson2010antagonistic,hillesland2010rapid}
on the way in which antagonistic coevolution can accelerate evolution
and dominate a systems's response to changing environmental conditions.
However, this topic has received much less theoretical attention to
date.  To investigate the interaction of genomic evolution and
population dynamics, we require that the matrix describing the
interactions between predator and prey evolve with genomic
fitness\cite{KOND03,ACKL04}.  In these approaches, the interaction
matrix is a linear function of either the fitness of the predators or
the relative fitness of a pair, which is appropriate to describe the
stabilization of complex foodwebs with a large number of species, but
probably not adequate to describe systems with a Red Queen dynamics,
where the genomic fitness is a distribution for a species instead of a
single number. In this case, the evolution of the interaction matrix is
nonlinear, and the dynamics needs to be treated as a stochastic
individual-level model\cite{WANG10}.

Coevolutionary dynamics is also an important arena for game theory
dynamics to play out in evolution and ecology, because the level of
cooperation between organisms can be analyzed on an
encounter-by-encounter basis as the repetition of a cooperative game,
such as the Prisoner's
Dilemma\cite{AXEL81,SMIT82,NOWA04,NOWA06,MCGI07}.  In the Prisoner's
Dilemma, the two players have two states: they can either \lq\lq
cooperate" or \lq\lq defect". If they cooperate, they receive a reward
$R$, if they both defect, they receive a punishment $P$, and if one
defects while the other cooperates, the former receives a temptation
$T$ while the latter receives the sucker's reward $S$.  If $T>R>P>S$,
then the following dilemma arises: a rational player would defect,
because it yields the highest reward independent of the state of the
other player.  So in a contest with two rational players, each will end
up with the punishment $P$, which is a shame: because if they had both
cooperated, they would have received the reward $R$.  This game
illustrates the paradoxical nature of two-body interactions in
cooperative dynamical systems, but can sometimes provide an accurate
idealization of actual biological interactions, if the biology can be
meaningfully mapped into simple game theory interactions terms.

In a remarkable experiment, Turner and Chao\cite{TURN99} studied the
evolution of fitness (measured in terms of population growth) of phages
that can multiply infect the same bacterial cell. Viruses can cooperate
by sharing intracellular enzymes needed for reproduction and can defect
by sequestering the enzymes. Turner and Chao engineered two strains to
behave as cooperators and defectors, and found that the strain with
high rate of co-infection initially increased in fitness over time, but
eventually evolved lower fitness, a counter-example to the usual
cavalier assumption that \lq\lq fitness always increases".  In this
case, Turner and Chao were able to show that this decline in fitness
arose from collective effects: the fitness of the virus strains
conformed to the Prisoner's Dilemma, whose pay-off matrix they were
able to measure.  Game theory is not the only way to interpret this
finding\cite{BROW01}, but the key point is that collective effects
provide important and sometimes counter-intuitive influences in
ecological and evolutionary interactions. The huge literature on this
area is beyond the scope of this review, but other ecological
interactions are measurable and interpretable in game theory
terms\cite{KIRK04,GORE09}.

We have emphasized in this review that the essence of evolution is
self-reference, but it is apparent that this is not captured by the
game theory models described above.  The problem is that the pay-off
matrix is given a priori, and is not able to evolve along with
the system itself.  Thus, the effect of each agents' actions on
changing the behavior of other players and the dynamics of the
environment is neglected.  A bona fide game theory approach to
evolution would allow the game rules themselves to change as a function
of the state of the players and their intrinsic dynamics.  This is
important for the following reason: in the usual formulation of game
theory\cite{SMIT82}, there is an equilibrium state that can arise known
as an evolutionary stable strategy (ESS).  Loosely speaking, this can
arise when a population of individuals play in a cooperative game in
which the payoff represents the fitness, and selection is
assumed to be operating.  An ESS is a strategy that if used by a large
enough number of individuals in a population is resistant to invasion
by alternative strategies, and as such represents some sort of
equilibrium (in fact a Nash equilibrium\cite{NASH50}). This equilibrium
state is in some sense analogous to thermal equilibrium and reflects
the static nature of the game itself.  If the game was allowed instead
to be dynamic, with the rules of the game able to change due to the
states of the players, then in addition to static equilibria there
could also be dynamic equilibria, analogous to a non-equilibrium steady
state.  Such a game could describe a steadily evolving system, one with
a stationary complexity cascade (itself is analogous to a turbulent
non-equilibrium state).

Akiyama and Kaneko\cite{AKIY2000,AKIY02} seem to be the only
researchers who have explicitly argued along these lines, making a
first step towards a theory for truly-evolving games using dynamical
systems theory to analyze a cooperative game with an evolving pay-off
matrix. Other game theoretic models that include some sort of dynamical
or learning behavior, such as players using their scores to adjust the
frequency of a finite set of a priori given responses, have been
developed, describing the system trajectories, showing that chaotic
dynamics can arise, and shedding doubt about the applicability of Nash
equilibria in the real world of dynamical games\cite{SATO02}---and by
extension to evolution.

A related development has been the dynamics of spatially-extended game
theoretic models that study the evolution of cooperation, in which
there is coevolution of the network of connections between
players\cite{ZIMM01,EBEL02} and the players' strategies as the game
progresses.  Typically players update their strategy based upon their
interactions with their network neighbours during the game,
representing perhaps the simplest game theory models where there is
feedback between the environment and the agents\cite{PERC09}.  Finally,
there has been progress in the most difficult aspect of the
relationship between game theory and evolutionary biology:
understanding how effective game theoretic description can arise from
the macroscopic dynamics of agents interacting with their
environment\cite{SATO05}.  We conclude this section by recalling that
evolutionary dynamics is more general than biology itself, and thus it
should come as no surprise that there are applications of evolutionary
game theory to evolutionary finance\cite{FARM99,HENS05}, and indeed
meaningful analogies between ecosystem dynamics and
finance\cite{MAY08}.

\section{Is evolution random?}

We would be remiss in ending this article if we did not briefly mention
the fascinating question: is evolution random? More
precisely, does variation precede but not cause adaptation---the
central tenet of the modern synthesis---or do environmental changes
alter the stochastic nature of the evolutionary process?  Any
indication that organisms can chose which mutations arise after an
environmental stress has been applied would be anathema to the central
tenet of the modern synthesis, and would require a re-evaluation of how
evolution is widely understood.

In a classic experiment involving the exposure of a strain of {\it E.
coli\/}
to bacteriophages, Luria and Delbr\"uck\cite{LURI43} showed that the
probability distribution for the number of mutants exhibited the
characteristics expected only if random mutations had been present
before exposure to the phage, and apparently ruling out the hypothesis
that mutations occurred as a result of the phage.  This might seem to
put the matter to rest, but because there is a priori no theoretical
reason why a cell could not sense environmental stress and respond in a
non-random way, researchers have persisted in exploring this issue
experimentally.  Although early experiments are generally recognized as
not being properly analyzed\cite{SNIE95}, a plethora of mechanisms have
now been reported to give rise to an adaptive response to stress,
including regulation of mutation rates (non-random in time) and
localized variation along the genome (non-random in genome
space)\cite{caporale2003darwin,GALH07,RAND07}.

There is also compelling evidence that not only may mutations be
non-random but horizontal gene transfer too need not be random.
Enterococcus faecalis, a gut-dwelling bacterium, can be resistant to
certain antibiotics if it contains the plasmid (an extrachromosmal loop
of DNA) pCF10.  This plasmid can be horizontally transferred from a
donor with the plasmid to a recipient initially without it, through the
process of conjugation (bacterial sex).  The remarkable feature of this
organism, however, is that the transfer is controlled by and initiated
by signals sent from the recipient\cite{dunny2007peptide}.  The
vancomycin resistant strain V583 of this organism is now one of the
leading causes of hospital-acquired infection, spreading rapidly
through horizontal gene transfer\cite{MANS10}.

\section{Conclusion}

In the natural development of the sciences, issues of complexity are
sensibly postponed until they can no longer be avoided. Physics was
able to delay serious consideration of collective effects for nearly
three hundred years, and only in the last thirty years or so has it
confronted complex collective phenomena involving multiple scales of
space and time, unpredictable dynamics and large fluctuations.  Its
track record of success is mixed.

Biology was not so lucky: at its outset, complex phenomena were
encountered, but tools were lacking to cope with the difficulty.
Rather than abiding by ignorance, a language-culture was developed to
explain away the conceptual difficulties using guesswork solutions such
as \lq\lq natural selection".  As Schr\"odinger wrote,
\begin{extract}
\lq\lq Instead of filling a gap by guesswork, genuine science prefers to put
up with it; and this, not so much from conscientious scruples about
telling lies, as from the consideration that, however irksome
the gap may be, its obliteration by a fake removes the urge to seek
after a tenable answer."---E. Schr\"odinger, Nature and the Greeks,
pp7-8.\cite{SCHRO54}
\end{extract}
Today, with the \lq\lq urge" removed, the development of sophisticated
technology has allowed biology to take refuge in single-molecule
biophysics, genomics and molecular biology.  But the stultifying
language-culture still remains. This sanctuary is an illusionary
respite: the core problems of biology remain irksome to some, and are
inextricably interwoven with evolution.  Indeed, the very existence of
biological phenomena is an expression of physical laws that represent a
new asymptotic realm in nonequilibrium statistical physics.  Ulam
famously quipped\cite{FRAU99} \lq\lq Ask not what physics can do for
biology; ask what biology can do for physics." Our answer is clear.

\section{Acknowledgements}


We thank the following for their helpful comments on earlier versions
of this manuscript: Elbert Branscombe, Lynn Caporale, Jim Davis,
Michael Deem, Barbara Drossel, Freeman Dyson, George Dyson, Larry Gold,
Leo Kadanoff, Alan McKane, Tim Newman, Jim Shapiro, and especially James Langer and
Kalin Vetsigian. This work was partially supported by the US National
Science Foundation under Grants NSF-EF-0526747 and NSF-PFC-082265,
through the Center for the Physics of Living Cells.

\bibliographystyle{ieeetr}
\bibliography{life}

\begin{thebibliography}{100}

\bibitem{lander2001initial}
E.~Lander, L.~Linton, B.~Birren, C.~Nusbaum, M.~Zody, J.~Baldwin, K.~Devon,
  K.~Dewar, M.~Doyle, W.~FitzHugh, {\em et~al.}, ``{Initial sequencing and
  analysis of the human genome},'' {\em Nature}, vol.~409, no.~6822,
  pp.~860--921, 2001.

\bibitem{CLAM07}
M.~Clamp, B.~Fry, M.~Kamal, X.~Xie, J.~Cuff, M.~F. Lin, M.~Kellis,
  K.~Lindblad-Toh, and E.~S. Lander, ``{Distinguishing protein-coding and
  noncoding genes in the human genome},'' {\em Proceedings of the National
  Academy of Sciences}, vol.~104, no.~49, pp.~19428--19433, 2007.

\bibitem{GOUL01}
S.~J. Gould, ``Humbled by the genome's mysteries.'' New York Times, Feb 21
  2001.

\bibitem{ochman2000lgt}
H.~Ochman, J.~Lawrence, E.~Groisman, {\em et~al.}, ``{Lateral gene transfer and
  the nature of bacterial innovation},'' {\em Nature}, vol.~405, no.~6784,
  pp.~299--304, 2000.

\bibitem{BASS05}
C.~Waters and B.~Bassler, ``{Quorum Sensing: Cell-to-Cell Communication in
  Bacteria},'' {\em Annu. Rev. Cell Dev. Biol}, vol.~21, pp.~319--46, 2005.

\bibitem{HARS03}
R.~Harshey, ``{Bacterial Motility On A Surface: Many Ways To A Common Goal},''
  {\em Annual Reviews in Microbiology}, vol.~57, no.~1, pp.~249--273, 2003.

\bibitem{COPE09}
M.~Copeland and D.~Weibel, ``{Bacterial swarming: a model system for studying
  dynamic self-assembly},'' {\em Soft Matter}, vol.~5, no.~6, pp.~1174--1187,
  2009.

\bibitem{SHAP98}
J.~Shapiro, ``{Thinking about bacterial populations as multicellular
  organisms},'' {\em Annual Reviews in Microbiology}, vol.~52, no.~1,
  pp.~81--104, 1998.

\bibitem{WOES09}
C.~Woese and N.~Goldenfeld, ``{How the microbial world saved evolution from the
  scylla of molecular biology and the charybdis of the modern synthesis},''
  {\em Microbiology and Molecular Biology Reviews}, vol.~73, no.~1, pp.~14--21,
  2009.

\bibitem{herz1994collective}
A.~Herz, ``{Collective phenomena in spatially extended evolutionary games},''
  {\em Journal of theoretical biology}, vol.~169, no.~1, pp.~65--87, 1994.

\bibitem{nowak2006evolutionary}
A.~Nowak~Martin, {\em {Evolutionary Dynamics: Exploring the Equations of
  Life}}.
\newblock Cambridge University Press, 2006.

\bibitem{GOLD07}
N.~Goldenfeld and C.~Woese, ``{Biology's next revolution},'' {\em Nature},
  vol.~445, p.~369, 2007.

\bibitem{shapiro2010mobile}
J.~Shapiro, ``{Mobile DNA and evolution in the 21st century.},'' {\em Mobile
  DNA}, vol.~1, no.~4, pp.~1--14, 2010.

\bibitem{levin1992problem}
S.~Levin, ``{The problem of pattern and scale in ecology: the Robert H.
  MacArthur award lecture},'' {\em Ecology}, vol.~73, no.~6, pp.~1943--1967,
  1992.

\bibitem{toner2005hydrodynamics}
J.~Toner, Y.~Tu, and S.~Ramaswamy, ``{Hydrodynamics and phases of flocks},''
  {\em Annals of Physics}, vol.~318, no.~1, pp.~170--244, 2005.

\bibitem{sumpter2006principles}
D.~Sumpter, ``{The principles of collective animal behaviour},'' {\em
  Philosophical Transactions of the Royal Society B: Biological Sciences},
  vol.~361, no.~1465, pp.~5--22, 2006.

\bibitem{schweitzer2007brownian}
F.~Schweitzer and J.~Farmer, {\em {Brownian Agents and Active Particles:
  Collective Dynamics in the Natural and Social Sciences}}.
\newblock Springer Verlag, 2007.

\bibitem{HALO08}
J.~Haloin and S.~Strauss, ``{Interplay between Ecological Communities and
  Evolution},'' {\em Annals of the New York Academy of Sciences}, vol.~1133,
  no.~The Year in Evolutionary Biology 2008, pp.~87--125, 2008.

\bibitem{cohen2007explaining}
I.~Cohen and D.~Harel, ``{Explaining a complex living system: dynamics,
  multi-scaling and emergence},'' {\em Journal of the Royal Society Interface},
  vol.~4, no.~13, pp.~175--182, 2007.

\bibitem{cohen2007real}
I.~Cohen, ``{Real and artificial immune systems: computing the state of the
  body},'' {\em Nature Reviews Immunology}, vol.~7, no.~7, pp.~569--574, 2007.

\bibitem{ben2000cooperative}
E.~Ben-Jacob, I.~Cohen, and H.~Levine, ``{Cooperative self-organization of
  microorganisms},'' {\em Advances in Physics}, vol.~49, no.~4, pp.~395--554,
  2000.

\bibitem{WING06}
N.~Wingreen and S.~Levin, ``{Cooperation among microorganisms},'' {\em PLoS
  Biology}, vol.~4, no.~9, pp.~1486--1488, 2006.

\bibitem{velicer2009evolution}
G.~Velicer and M.~Vos, ``{Sociobiology of the myxobacteria},'' {\em Annual
  review of microbiology}, vol.~63, pp.~599--623, 2009.

\bibitem{LENT08}
T.~Lenton, H.~Held, E.~Kriegler, J.~Hall, W.~Lucht, S.~Rahmstorf, and
  H.~Schellnhuber, ``{Tipping elements in the Earth's climate system},'' {\em
  Proceedings of the National Academy of Sciences}, vol.~105, no.~6,
  pp.~1786--1793, 2008.

\bibitem{BARD08}
R.~Bardgett, C.~Freeman, and N.~Ostle, ``{Microbial contributions to climate
  change through carbon cycle feedbacks},'' {\em The ISME Journal}, vol.~2,
  no.~8, pp.~805--814, 2008.

\bibitem{HAST09}
K.~Cuddington, W.~Wilson, and A.~Hastings, ``{Ecosystem Engineers: Feedback and
  Population Dynamics},'' {\em The American Naturalist}, vol.~173, no.~4,
  pp.~488--498, 2009.

\bibitem{SMIT93}
J.~Maynard~Smith, ``{The theory of evolution},'' {\em CUP, Cambridge}, 1993.

\bibitem{DROS01}
B.~Drossel, ``{Biological evolution and statistical physics},'' {\em Advances
  in Physics}, vol.~50, no.~2, pp.~209--295, 2001.

\bibitem{SAPP09}
J.~Sapp, {\em {The new foundations of evolution: on the tree of life}}.
\newblock Oxford University Press, USA, 2009.

\bibitem{salyers1997arg}
A.~Salyers and C.~Amabile-Cuevas, ``{Why are antibiotic resistance genes so
  resistant to elimination?},'' {\em Antimicrobial agents and chemotherapy},
  vol.~41, pp.~2321--2325, 1997.

\bibitem{gatenby2009change}
R.~Gatenby, ``{A change of strategy in the war on cancer},'' {\em Nature},
  vol.~459, no.~7246, pp.~508--509, 2009.

\bibitem{LANG80}
J.~Langer, ``{Instabilities and pattern formation in crystal growth},'' {\em
  Reviews of Modern Physics}, vol.~52, no.~1, pp.~1--28, 1980.

\bibitem{VANH48}
J.~B.~V. Helmont, {\em Ortus Medicinae}.
\newblock Amsterdam, 1648.
\newblock English translation in Great Experiments in Biology, p. 155, Gabriel,
  M.L. and Fogel, S. (eds.) (Prentice Hall, NJ, 1955).

\bibitem{EINS22}
A.~Einstein, {\em Theoretische Bemerkungen zur Supraleitung der Metalle}.
\newblock Eduardo Ijdo, Leiden, 1922.
\newblock [English translation by Bjoern Schmekel as Theoretical remark on the
  superconductivity of metals, available online at
  http://arxiv.org/abs/physics/0510251].

\bibitem{sauer2007einstein}
T.~Sauer, ``{Einstein and the Early Theory of Superconductivity, 1919--1922},''
  {\em Archive for history of exact sciences}, vol.~61, no.~2, pp.~159--211,
  2007.

\bibitem{feynman1953atomic}
R.~Feynman, ``{Atomic theory of the $\lambda$ transition in Helium},'' {\em
  Physical Review}, vol.~91, no.~6, pp.~1291--1301, 1953.

\bibitem{ALFO08}
M.~Alford, A.~Schmitt, K.~Rajagopal, and T.~Sch{\\"a}fer, ``{Color
  superconductivity in dense quark matter},'' {\em Reviews of Modern Physics},
  vol.~80, no.~4, pp.~1455--1515, 2008.

\bibitem{gladyshev2008massive}
E.~Gladyshev, M.~Meselson, and I.~Arkhipova, ``{Massive horizontal gene
  transfer in bdelloid rotifers},'' {\em Science}, vol.~320, no.~5880, p.~1210,
  2008.

\bibitem{keeling2008horizontal}
P.~Keeling and J.~Palmer, ``{Horizontal gene transfer in eukaryotic
  evolution},'' {\em Nature Reviews Genetics}, vol.~9, no.~8, pp.~605--618,
  2008.

\bibitem{palenik2009coastal}
B.~Palenik, Q.~Ren, V.~Tai, and I.~Paulsen, ``{Coastal Synechococcus metagenome
  reveals major roles for horizontal gene transfer and plasmids in population
  diversity},'' {\em Environmental Microbiology}, vol.~11, no.~2, pp.~349--359,
  2009.

\bibitem{monier2009horizontal}
A.~Monier, A.~Pagarete, C.~de~Vargas, M.~Allen, B.~Read, J.~Claverie, and
  H.~Ogata, ``{Horizontal gene transfer of an entire metabolic pathway between
  a eukaryotic alga and its DNA virus},'' {\em Genome Research}, vol.~19,
  no.~8, p.~1441, 2009.

\bibitem{pace2008repeated}
J.~Pace, C.~Gilbert, M.~Clark, and C.~Feschotte, ``{Repeated horizontal
  transfer of a DNA transposon in mammals and other tetrapods},'' {\em
  Proceedings of the National Academy of Sciences}, vol.~105, no.~44, p.~17023,
  2008.

\bibitem{DANO10}
R.~Danovaro, A.~Dell'Anno, A.~Pusceddu, C.~Gambi, I.~Heiner, and
  R.~Mobjerg~Kristensen, ``The first metazoa living in permanently anoxic
  conditions,'' {\em BMC Biology}, vol.~8, no.~1, p.~30, 2010.

\bibitem{kimmel2010evolution}
M.~Kimmel, ``{Evolution and cancer: a mathematical biology approach},'' {\em
  Biology Direct}, vol.~5, no.~1, p.~29, 2010.

\bibitem{attolini2009evolutionary}
C.~Attolini and F.~Michor, ``{Evolutionary theory of cancer},'' {\em Annals of
  the New York Academy of Sciences}, vol.~1168, no.~1, pp.~23--51, 2009.

\bibitem{DUFF08}
S.~Duffy, L.~Shackelton, and E.~Holmes, ``{Rates of evolutionary change in
  viruses: patterns and determinants},'' {\em Nature Reviews Genetics}, vol.~9,
  no.~4, pp.~267--276, 2008.

\bibitem{NEHE10}
R.~Neher and T.~Leitner, ``{Recombination rate and selection strength in HIV
  intra-patient evolution},'' {\em PLoS Computational Biology}, vol.~6, no.~1,
  pp.~e10006600/1--7, 2010.

\bibitem{FALK08}
P.~G. Falkowski and L.~V. Godfrey, ``{Electrons, life and the evolution of
  Earth's oxygen cycle},'' {\em Philosophical Transactions of the Royal Society
  B: Biological Sciences}, vol.~363, no.~1504, pp.~2705--2716, 2008.

\bibitem{DELB49}
M.~Delbr\"uck, ``A physicist looks at biology,'' {\em Trans. Conn. Acad. Arts
  and Sciences.}, vol.~38, pp.~173--190, 1949.

\bibitem{wang98}
J.~Wang, ``{Moving one DNA double helix through another by a type II DNA
  topoisomerase: the story of a simple molecular machine},'' {\em Quarterly
  reviews of biophysics}, vol.~31, no.~02, pp.~107--144, 1998.

\bibitem{WANG09}
J.~Wang, ``{A Journey in the World of DNA Rings and Beyond},'' {\em Annual
  review of biochemistry}, vol.~78, pp.~31--54, 2009.
\newblock See p.40,second column.

\bibitem{WALL1858}
A.~Wallace, ``{On the tendency of varieties to depart indefinitely from the
  original type},'' {\em Journal of the Linnean Society of London, Zoology},
  vol.~3, pp.~53--62, 1858.

\bibitem{DARW1859}
C.~Darwin, {\em {On the origin of species by means of natural selection, or the
  preservation of favoured races in the struggle for life (1859)}}.
\newblock J. Murray, London, UK, 1859.

\bibitem{HUXL42}
J.~Huxley, {\em {Evolution: the modern synthesis}}.
\newblock MIT Press, 1942.

\bibitem{KIMU85}
M.~Kimura, {\em {The neutral theory of molecular evolution}}.
\newblock Cambridge University Press, 1985.

\bibitem{wright1932rmi}
S.~Wright, ``{The roles of mutation, inbreeding, crossbreeding and selection in
  evolution},'' {\em Proceedings of the Sixth International Congress on
  Genetics}, vol.~1, no.~6, pp.~356--366, 1932.

\bibitem{gavrilets2004fla}
S.~Gavrilets, {\em {Fitness landscapes and the origin of species}}.
\newblock Princeton University Press, Princeton, US, 2004.

\bibitem{orr2005gta}
H.~Orr, ``{The genetic theory of adaptation: a brief history},'' {\em Nature
  Reviews Genetics}, vol.~6, no.~2, pp.~119--127, 2005.

\bibitem{PARK10}
S.~Park, D.~Simon, and J.~Krug, ``{The speed of evolution in large asexual
  populations},'' {\em Journal of Statistical Physics}, vol.~138, no.~1,
  pp.~381--410, 2010.

\bibitem{NEHE09}
R.~Neher and B.~Shraiman, ``{Competition between recombination and epistasis
  can cause a transition from allele to genotype selection},'' {\em Proceedings
  of the National Academy of Sciences}, vol.~106, no.~16, pp.~6866--6871, 2009.

\bibitem{GING09}
P.~Gingerich, ``{Rates of Evolution},'' {\em Annual Review of Ecology,
  Evolution, and Systematics}, vol.~40, pp.~657--675, 2009.

\bibitem{FISH07}
D.~Fisher, ``{Course 11: Evolutionary dynamics},'' in {\em Les Houches Summer
  School Proceedings}, vol.~85, pp.~395--446, 2007.

\bibitem{desai2007speed}
M.~Desai, D.~Fisher, and A.~Murray, ``{The speed of evolution and maintenance
  of variation in asexual populations},'' {\em Current Biology}, vol.~17,
  no.~5, pp.~385--394, 2007.

\bibitem{brunet2008stochastic}
E.~Brunet, I.~Rouzine, and C.~Wilke, ``{The stochastic edge in adaptive
  evolution.},'' {\em Genetics}, vol.~179, no.~1, pp.~603--620, 2008.

\bibitem{fogle2008clonal}
C.~Fogle, J.~Nagle, and M.~Desai, ``{Clonal Interference, Multiple Mutations
  and Adaptation in Large Asexual Populations},'' {\em Genetics}, vol.~180,
  no.~4, pp.~2163--2173, 2008.

\bibitem{weissman2009rate}
D.~Weissman, M.~Desai, D.~Fisher, and M.~Feldman, ``{The rate at which asexual
  populations cross fitness valleys},'' {\em Theoretical population biology},
  vol.~75, no.~4, pp.~286--300, 2009.

\bibitem{MOOR05}
J.~Moore, ``{A global view of epistasis},'' {\em Nature genetics}, vol.~37,
  no.~1, pp.~13--14, 2005.

\bibitem{SEGR05}
D.~Segre, A.~DeLuna, G.~Church, and R.~Kishony, ``{Modular epistasis in yeast
  metabolism},'' {\em Nature genetics}, vol.~37, no.~1, pp.~77--83, 2005.

\bibitem{VETS05}
K.~Vetsigian and N.~Goldenfeld, ``{Global divergence of microbial genome
  sequences mediated by propagating fronts},'' {\em Proceedings of the National
  Academy of Sciences}, vol.~102, no.~20, pp.~7332--7337, 2005.

\bibitem{CHEN09}
J.~Chen, Y.~Wu, H.~Yang, J.~Bergelson, and M.~Kreitman, ``{Variation in the
  Ratio of Nucleotide Substitution and Indel Rates across Genomes in Mammals
  and Bacteria},'' {\em Molecular Biology and Evolution}, vol.~26, no.~7,
  pp.~1523--1531, 2009.

\bibitem{MCCL84}
B.~McClintock, ``{The Significance of Responses of the Genome to Challenge},''
  {\em Science}, vol.~226, pp.~792--801, 1984.

\bibitem{SHAP09}
J.~Shapiro, ``{Revisiting the central dogma in the 21st Century},'' {\em Annals
  of the New York Academy of Sciences}, vol.~1178, pp.~6--28, 2009.

\bibitem{syvanen1994hgt}
M.~Syvanen, ``{Horizontal gene transfer: evidence and possible consequences},''
  {\em Annual Review of Genetics}, vol.~28, no.~1, pp.~237--261, 1994.

\bibitem{filee2003rpv}
J.~Filee, P.~Forterre, and J.~Laurent, ``{The role played by viruses in the
  evolution of their hosts: a view based on informational protein
  phylogenies},'' {\em Res. Microbiol}, vol.~154, pp.~237--243, 2003.

\bibitem{weinbauer2004vdm}
M.~Weinbauer and F.~Rassoulzadegan, ``{Are viruses driving microbial
  diversification and diversity?},'' {\em Environmental Microbiology}, vol.~6,
  no.~1, pp.~1--11, 2004.

\bibitem{frost2005mobile}
L.~Frost, R.~Leplae, A.~Summers, and A.~Toussaint, ``{Mobile genetic elements:
  the agents of open source evolution},'' {\em Nature Reviews Microbiology},
  vol.~3, no.~9, pp.~722--732, 2005.

\bibitem{babic2008direct}
A.~Babic, A.~Lindner, M.~Vulic, E.~Stewart, and M.~Radman, ``{Direct
  visualization of horizontal gene transfer},'' {\em Science}, vol.~319,
  no.~5869, pp.~1533--1536, 2008.

\bibitem{PAL07}
C.~Pal, M.~Macia, A.~Oliver, I.~Schachar, and A.~Buckling, ``Coevolution with
  viruses drives the evolution of bacterial mutation rates,'' {\em Nature},
  vol.~450, pp.~1079--1081, 2007.

\bibitem{PAUL10}
L.~D. McDaniel, E.~Young, J.~Delaney, F.~Ruhnau, K.~B. Ritchie, and J.~H. Paul,
  ``High frequency of horizontal gene transfer in the oceans,'' {\em Science},
  vol.~330, p.~50, 2010.

\bibitem{HOTOPP07}
J.~Hotopp, M.~Clark, D.~Oliveira, J.~Foster, P.~Fischer, M.~Torres, J.~Giebel,
  N.~Kumar, N.~Ishmael, S.~Wang, {\em et~al.}, ``{Widespread lateral gene
  transfer from intracellular bacteria to multicellular eukaryotes},'' {\em
  Science}, vol.~317, no.~5845, pp.~1753--1756, 2007.

\bibitem{biémont2006genetics}
C.~Bi{\'e}mont and C.~Vieira, ``{Genetics: junk DNA as an evolutionary
  force},'' {\em Nature}, vol.~443, no.~7111, pp.~521--524, 2006.

\bibitem{schaack2010promiscuous}
S.~Schaack, C.~Gilbert, and C.~Feschotte, ``{Promiscuous DNA: horizontal
  transfer of transposable elements and why it matters for eukaryotic
  evolution},'' {\em Trends in Ecology \& Evolution}, 2010.
\newblock in press.

\bibitem{rankin2010traits}
D.~Rankin, E.~Rocha, and S.~Brown, ``{What traits are carried on mobile genetic
  elements, and why\&quest},'' {\em Heredity}, vol.~1, pp.~1--10, 2010.
\newblock in press.

\bibitem{venner2009dynamics}
S.~Venner, C.~Feschotte, and C.~Bi{\'e}mont, ``{Dynamics of transposable
  elements: towards a community ecology of the genome},'' {\em Trends in
  Genetics}, vol.~25, no.~7, pp.~317--323, 2009.

\bibitem{LAMA1809}
J.~de~Lamarck, {\em {Philosophie zoologique}}.
\newblock C. Martins, 1809.

\bibitem{KOON09}
E.~Koonin and Y.~Wolf, ``{Is evolution Darwinian or/and Lamarckian?},'' {\em
  Biology Direct}, vol.~4, no.~42, pp.~1--14, 2009.

\bibitem{JABL98}
E.~Jablonka and M.~Lamb, ``{Epigenetic inheritance in evolution},'' {\em
  Journal of Evolutionary Biology}, vol.~11, no.~2, pp.~159--183, 1998.

\bibitem{RAND07}
O.~Rando and K.~Verstrepen, ``{Timescales of genetic and epigenetic
  inheritance},'' {\em Cell}, vol.~128, no.~4, pp.~655--668, 2007.

\bibitem{NOWA09}
M.~Nowacki and L.~Landweber, ``{Epigenetic inheritance in ciliates},'' {\em
  Current opinion in microbiology}, vol.~12, no.~6, pp.~638--643, 2009.

\bibitem{GUTT08}
N.~Guttenberg and N.~Goldenfeld, ``{Cascade of Complexity in Evolving
  Predator-Prey Dynamics},'' {\em Physical Review Letters}, vol.~100, no.~5,
  p.~58102, 2008.

\bibitem{caporale2003darwin}
L.~Caporale, {\em {Darwin in the genome: molecular strategies in biological
  evolution}}.
\newblock McGraw-Hill Professional, 2003.

\bibitem{HAIR05}
N.~G. Hairston, S.~P. Ellner, M.~A. Geber, T.~Yoshida, and F.~J. A, ``Rapid
  evolution and the convergence of ecological and evolutionary time,'' {\em
  Ecology Letters}, vol.~8, pp.~1114--1127, 2005.

\bibitem{cowen2005hsp90}
L.~Cowen and S.~Lindquist, ``{Hsp90 potentiates the rapid evolution of new
  traits: drug resistance in diverse fungi},'' {\em Science}, vol.~309,
  no.~5744, p.~2185, 2005.

\bibitem{thompson1998rapid}
J.~Thompson, ``{Rapid evolution as an ecological process},'' {\em Trends in
  Ecology \& Evolution}, vol.~13, no.~8, pp.~329--332, 1998.

\bibitem{carroll2007evolution}
S.~Carroll, A.~Hendry, D.~Reznick, and C.~Fox, ``{Evolution on ecological
  time-scales},'' {\em Ecology}, vol.~21, pp.~387--393, 2007.

\bibitem{steinhauer1987rapid}
D.~Steinhauer and J.~Holland, ``{Rapid evolution of RNA viruses},'' {\em Annual
  Reviews in Microbiology}, vol.~41, no.~1, pp.~409--431, 1987.

\bibitem{zhang2008rapid}
Z.~Zhang, G.~Weinstock, and M.~Gerstein, ``{Rapid evolution by positive
  Darwinian selection in T-cell antigen CD4 in primates},'' {\em Journal of
  Molecular Evolution}, vol.~66, no.~5, pp.~446--456, 2008.

\bibitem{summers2008molecular}
K.~Summers and B.~Crespi, ``{Molecular evolution of the prostate cancer
  susceptibility locus RNASEL: evidence for positive selection},'' {\em
  Infection, Genetics and Evolution}, vol.~8, no.~3, pp.~297--301, 2008.

\bibitem{meyer2006prey}
J.~Meyer, S.~Ellner, N.~Hairston, L.~Jones, and T.~Yoshida, ``{Prey evolution
  on the time scale of predator--prey dynamics revealed by allele-specific
  quantitative PCR},'' {\em Proceedings of the National Academy of Sciences},
  vol.~103, no.~28, pp.~10690--10695, 2006.

\bibitem{hillesland2009experimental}
K.~Hillesland, G.~Velicer, and R.~Lenski, ``{Experimental evolution of a
  microbial predator's ability to find prey},'' {\em Proceedings of the Royal
  Society B: Biological Sciences}, vol.~276, no.~1656, pp.~459--467, 2009.

\bibitem{palkovacs2009experimental}
E.~Palkovacs and D.~Post, ``{Experimental evidence that phenotypic divergence
  in predators drives community divergence in prey},'' {\em Ecology}, vol.~90,
  no.~2, pp.~300--305, 2009.

\bibitem{bailey2009genes}
J.~Bailey, A.~Hendry, M.~Kinnison, D.~Post, E.~Palkovacs, F.~Pelletier,
  L.~Harmon, and J.~Schweitzer, ``{From genes to ecosystems: an emerging
  synthesis of eco-evolutionary dynamics},'' {\em New Phytologist}, vol.~184,
  no.~4, pp.~746--749, 2009.

\bibitem{pelletier2009eco}
F.~Pelletier, D.~Garant, and A.~Hendry, ``{Eco-evolutionary dynamics},'' {\em
  Philosophical Transactions of the Royal Society B: Biological Sciences},
  vol.~364, no.~1523, pp.~1483--1489, 2009.

\bibitem{jones2009rapid}
L.~Jones, L.~Becks, S.~Ellner, N.~Hairston~Jr, T.~Yoshida, and G.~Fussmann,
  ``{Rapid contemporary evolution and clonal food web dynamics},'' {\em
  Philosophical Transactions of the Royal Society of London, Series B:
  Biological Sciences}, vol.~364, no.~1523, pp.~1579--1591, 2009.

\bibitem{terhorst2010evolution}
C.~terHorst, T.~Miller, and D.~Levitan, ``{Evolution of prey in ecological time
  reduces the effect size of predators in experimental microcosms.},'' {\em
  Ecology}, vol.~91, no.~3, pp.~629--636, 2010.

\bibitem{yoshida2007cryptic}
T.~Yoshida, S.~Ellner, L.~Jones, B.~Bohannan, R.~Lenski, and N.~Hairston~Jr,
  ``{Cryptic population dynamics: rapid evolution masks trophic
  interactions},'' {\em PLoS Biol}, vol.~5, no.~9, p.~e235, 2007.

\bibitem{BULM87}
M.~Bulmer, ``{Coevolution of codon usage and transfer RNA abundance},'' {\em
  Nature}, vol.~325, no.~6106, pp.~728--730, 1987.

\bibitem{BULM91}
M.~Bulmer, ``{The selection-mutation-drift theory of synonymous codon usage},''
  {\em Genetics}, vol.~129, no.~3, pp.~897--907, 1991.

\bibitem{VETS09}
K.~Vetsigian and N.~Goldenfeld, ``{Genome rhetoric and the emergence of
  compositional bias},'' {\em Proceedings of the National Academy of Sciences},
  vol.~106, no.~1, pp.~215--220, 2009.

\bibitem{HART99}
L.~Hartwell, J.~Hopfield, S.~Leibler, and A.~Murray, ``{From molecular to
  modular cell biology},'' {\em Nature}, vol.~402, pp.~C47--C52, 1999.

\bibitem{BARA04}
A.~Barab{\'a}si and Z.~Oltvai, ``{Network biology: understanding the cell's
  functional organization},'' {\em Nature Reviews Genetics}, vol.~5, no.~2,
  pp.~101--113, 2004.

\bibitem{KASH05}
N.~Kashtan and U.~Alon, ``{Spontaneous evolution of modularity and network
  motifs},'' {\em Proceedings of the National Academy of Sciences}, vol.~102,
  no.~39, pp.~13773--13778, 2005.

\bibitem{KASH06}
N.~Kashtan, E.~Noor, and U.~Alon, ``Varying environments can speed up
  evolution,'' {\em Proc.\ Natl.\ Acad.\ Sci.\ USA}, vol.~104,
  pp.~13711--13716, 2006.

\bibitem{HOLL92}
J.~Holland, {\em {Adaptation in natural and artificial systems}}.
\newblock MIT press Cambridge, MA, 1992.

\bibitem{EARL04}
D.~J. Earl and M.~W. Deem, ``Evolvability is a selectable trait,'' {\em Proc.\
  Natl.\ Acad.\ Sci.\ USA}, vol.~110, no.~32, pp.~11531--11536, 2004.

\bibitem{HE09}
J.~He, J.~Sun, and M.~Deem, ``{Spontaneous emergence of modularity in a model
  of evolving individuals and in real networks},'' {\em Physical Review E},
  vol.~79, no.~3, p.~31907, 2009.

\bibitem{PAL05}
C.~P{\'a}l, B.~Papp, and M.~Lercher, ``{Adaptive evolution of bacterial
  metabolic networks by horizontal gene transfer},'' {\em Nature genetics},
  vol.~37, no.~12, pp.~1372--1375, 2005.

\bibitem{PART07}
M.~Parter, N.~Kashtan, and U.~Alon, ``{Environmental variability and modularity
  of bacterial metabolic networks},'' {\em BMC Evolutionary Biology}, vol.~7,
  pp.~169/1--8, 2007.

\bibitem{KREI08}
A.~Kreimer, E.~Borenstein, U.~Gophna, and E.~Ruppin, ``{The evolution of
  modularity in bacterial metabolic networks},'' {\em Proceedings of the
  National Academy of Sciences}, vol.~105, no.~19, pp.~6976--6981, 2008.

\bibitem{sullivan2006pae}
M.~Sullivan, D.~Lindell, J.~Lee, L.~Thompson, J.~Bielawski, and S.~Chisholm,
  ``{Prevalence and Evolution of Core Photosystem II Genes in Marine
  Cyanobacterial Viruses and Their Hosts},'' {\em PLoS Biol}, vol.~4, no.~8,
  p.~e234, 2006.

\bibitem{ANDE66}
E.~S. Anderson, ``Possible importance of transfer factors in bacterial
  evolution,'' {\em Nature}, vol.~209, pp.~637--638, 1966.

\bibitem{ANDE70}
N.~G. Anderson, ``Evolutionary significance of virus infection,'' {\em Nature},
  vol.~227, pp.~1346--1347, 1970.

\bibitem{sonea1988bwl}
S.~Sonea, ``{A bacterial way of life},'' {\em Nature}, vol.~331, no.~6153,
  p.~216, 1988.

\bibitem{BERE09}
R.~Johnson, J.~Evans, G.~Robinson, and M.~Berenbaum, ``{Changes in transcript
  abundance relating to colony collapse disorder in honey bees (Apis
  mellifera)},'' {\em Proceedings of the National Academy of Sciences},
  vol.~106, no.~35, pp.~14790--14795, 2009.

\bibitem{HORI10}
M.~Horie, T.~Honda, Y.~Suzuki, Y.~Kobayashi, T.~Daito, T.~Oshida, K.~Ikuta,
  P.~Jern, T.~Gojobori, J.~Coffin, {\em et~al.}, ``{Endogenous non-retroviral
  RNA virus elements in mammalian genomes},'' {\em Nature}, vol.~463, no.~7277,
  pp.~84--87, 2010.

\bibitem{BELY10}
V.~Belyi, A.~Levine, and A.~Skalka, ``{Unexpected Inheritance: Multiple
  Integrations of Ancient Bornavirus and Ebolavirus/Marburgvirus Sequences in
  Vertebrate Genomes},'' {\em PLoS Pathog}, vol.~6, no.~7, pp.~e1001030/1--12,
  2010.

\bibitem{wallace1855law}
A.~Wallace, ``{On the law which has regulated the introduction of new
  species},'' {\em Annals and Magazine of Natural History}, vol.~16, no.~2,
  pp.~184--196, 1855.

\bibitem{ibrahim1996spatial}
K.~Ibrahim, R.~Nichols, and G.~Hewitt, ``{Spatial patterns of genetic variation
  generated by different forms of dispersal during range expansion},'' {\em
  Heredity}, vol.~77, no.~3, pp.~282--291, 1996.

\bibitem{hallatschek2008gene}
O.~Hallatschek and D.~Nelson, ``{Gene surfing in expanding populations.},''
  {\em Theoretical Population Biology}, vol.~73, no.~1, pp.~158--170, 2008.

\bibitem{hallatschek2009fisher}
O.~Hallatschek and K.~Korolev, ``{Fisher waves in the strong noise limit},''
  {\em Physical Review Letters}, vol.~103, no.~10, p.~108103, 2009.

\bibitem{korolev2009}
K.~Korolev, M.~Avlund, O.~Hallatschek, and D.~Nelson, ``{Genetic demixing and
  evolution in linear stepping stone models},'' {\em Reviews of Modern
  Physics}, vol.~82, pp.~1691--1718, 2010.

\bibitem{MCKA04}
A.~McKane and T.~Newman, ``{Stochastic models in population biology and their
  deterministic analogs},'' {\em Physical Review E}, vol.~70, no.~4, p.~41902,
  2004.

\bibitem{butler2009robust}
T.~Butler and N.~Goldenfeld, ``{Robust ecological pattern formation induced by
  demographic noise},'' {\em Physical Review E}, vol.~80, no.~3, p.~30902,
  2009.

\bibitem{ODLI03}
F.~Odling-Smee, K.~Laland, and M.~Feldman, {\em {Niche construction: the
  neglected process in evolution}}.
\newblock Princeton Univ Press, 2003.

\bibitem{DAY03}
R.~Day, K.~Laland, and J.~Odling-Smee, ``{Rethinking Adaptation},'' {\em
  Perspectives in Biology and Medicine}, vol.~46, no.~1, pp.~80--95, 2003.

\bibitem{LENS06}
R.~Lenski, J.~Barrick, and C.~Ofria, ``{Balancing robustness and
  evolvability},'' {\em PLoS Biol}, vol.~4, no.~e428, pp.~2190--2192, 2006.

\bibitem{PALU09}
S.~R. Palumbi, ``Better evolution through chemistry: evolution driven by human
  changes to the chemical environment,'' in {\em Chemical evolution II: from
  the origin of life to modern society} (L.~Zaikowski {\em et~al.}, eds.),
  American Chemical Society, 2009.

\bibitem{GOLDeditorial09}
N.~Goldenfeld, ``No man is an island: quoted in editorial on complex systems,''
  {\em Nature Physics}, vol.~5, p.~1, 2009.

\bibitem{RAY91}
T.~S. Ray, ``Evolution, ecology, and optimization of digital organisms,'' in
  {\em Scientific Excellence in Supercomputing: The IBM 1990 Contest Prize
  Papers} (K.~R. Billingsley, E.~Derohanes, and I.~H~Brown, eds.),
  pp.~489--531, The University of Georgia: The Baldwin Press, 1991.

\bibitem{ADAM94}
C.~Adami and C.~T. Brown, ``Evolutionary learning in the 2d artificial life
  systems: Avida,'' in {\em Proc. Artificial Life IV} (R.~Brooks and P.~Maes,
  eds.), (US), pp.~377--381, MIT Press, 1994.

\bibitem{CALD98}
G.~Caldarelli, P.~Higgs, and A.~McKane, ``{Modelling Coevolution in
  Multispecies Communities},'' {\em Journal of Theoretical Biology}, vol.~193,
  no.~2, pp.~345--358, 1998.

\bibitem{ADAM02}
C.~Adami, ``{What is complexity?},'' {\em BioEssays}, vol.~24, no.~12,
  pp.~1085--1094, 2002.

\bibitem{MCKA04a}
A.~McKane, ``{Evolving complex food webs},'' {\em The European Physical Journal
  B-Condensed Matter}, vol.~38, no.~2, pp.~287--295, 2004.

\bibitem{MARO06}
M.~Maron and C.~T. Fernando, ``Food webs and the evolution of organism
  complexity,'' in {\em Complexity Workshop, Artificial Life X}, 2006.

\bibitem{elena2003eem}
S.~Elena and R.~Lenski, ``{Evolution experiments with microorganisms: The
  dynamics and genetic bases of adaptation},'' {\em Nature Reviews: Genetics},
  vol.~4, no.~6, pp.~457--469, 2003.

\bibitem{ADAM06}
C.~Adami, ``{Digital genetics: unravelling the genetic basis of evolution},''
  {\em Nature Reviews Genetics}, vol.~7, pp.~109--118, 2006.

\bibitem{wilke2002bdo}
C.~Wilke and C.~Adami, ``{The biology of digital organisms},'' {\em Trends in
  Ecology \& Evolution}, vol.~17, no.~11, pp.~528--532, 2002.

\bibitem{LENS99}
R.~E. Lenski, C.~Ofria, T.~C. Collier, and C.~Adami, ``Genome complexity,
  robustness and genetic interactions in digital organisms,'' {\em Nature},
  vol.~400, pp.~661--664, 1999.

\bibitem{ADAM95}
C.~Adami, C.~T. Brown, and M.~R. Haggerty, ``Abundance-distributions in
  artificial life and stochastic models: ``age and area'' revisited,'' in {\em
  Proceedings of the Third European Conference on Advances in Artificial Life},
  (London, UK), pp.~503--514, Springer-Verlag, 1995.

\bibitem{ostrowski2007esa}
E.~Ostrowski, C.~Ofria, and R.~Lenski, ``{Ecological specialization and
  adaptive decay in digital organisms.},'' {\em Am. Nat}, vol.~169,
  pp.~E1--E20, 2007.

\bibitem{LENS03}
R.~Lenski, C.~Ofria, R.~T. Pennock, and C.~Adami, ``The evolutionary origin of
  complex features,'' {\em Nature}, vol.~423, pp.~139--144, 2003.

\bibitem{WOES71}
C.~Woese, ``{Evolution of macromolecular complexity},'' {\em Journal of
  Theoretical Biology}, vol.~33, no.~1, pp.~29--34, 1971.

\bibitem{JONE95}
S.~Jones and J.~Thornton, ``{Protein-protein interactions: a review of protein
  dimer structures.},'' {\em Progress in biophysics and molecular biology},
  vol.~63, no.~1, pp.~31--65, 1995.

\bibitem{GOOD00}
D.~Goodsell and A.~Olson, ``{Structural symmetry and protein function},'' {\em
  Annu. Rev. Biophys. Biomol. Struct}, vol.~29, pp.~105--153, 2000.

\bibitem{PLAX09}
K.~Plaxco and M.~Gross, ``{Protein Complexes: The Evolution of Symmetry},''
  {\em Current Biology}, vol.~19, no.~1, pp.~R25--R26, 2009.

\bibitem{PERE07}
J.~Pereira-Leal, E.~Levy, C.~Kamp, and S.~Teichmann, ``{Evolution of protein
  complexes by duplication of homomeric interactions},'' {\em Genome biology},
  vol.~8, no.~4, pp.~R51/1--12, 2007.

\bibitem{LEVY08}
E.~Levy, E.~Erba, C.~Robinson, and S.~Teichmann, ``{Assembly reflects evolution
  of protein complexes},'' {\em Nature}, vol.~453, no.~7199, pp.~1262--1265,
  2008.

\bibitem{richardson1922wpn}
L.~Richardson, {\em {Weather Prediction by Numerical Process}}.
\newblock Cambridge University Press, Cambridge, UK, 1922.

\bibitem{KOLM41}
A.~N. Kolmogorov, ``{Local structure of turbulence in an incompressible fluid
  at very high Reynolds numbers},'' {\em Dokl Acad Nauk USSR}, vol.~30,
  pp.~299--303, 1941.

\bibitem{kataoka2003dynamical}
N.~Kataoka and K.~Kaneko, ``{Dynamical networks in function dynamics},'' {\em
  Physica D: Nonlinear Phenomena}, vol.~181, no.~3-4, pp.~235--251, 2003.

\bibitem{ANDE09}
R.~Anderson and R.~May, ``{Coevolution of hosts and parasites},'' {\em
  Parasitology}, vol.~85, no.~02, pp.~411--426, 1982.

\bibitem{BASC07}
J.~Bascompte and P.~Jordano, ``{Plant-Animal Mutualistic Networks: The
  Architecture of Biodiversity},'' {\em Annu. Rev. Ecol. Evol. Syst}, vol.~38,
  pp.~567--93, 2007.

\bibitem{KIER08}
E.~Kiers and R.~Denison, ``{Sanctions, cooperation, and the stability of
  plant-rhizosphere mutualisms},'' {\em Annual review of ecology, evolution,
  and systematics}, vol.~39, pp.~215--236, 2008.

\bibitem{DETH07}
L.~Dethlefsen, M.~Mcfall-Nagai, and D.~A. Relman, ``{An ecological and
  evolutionary perspective on human-microbe mutualism and disease},'' {\em
  Nature}, vol.~449, no.~7164, pp.~811--818, 2007.

\bibitem{van1973new}
L.~Van~Valen, ``{A new evolutionary law},'' {\em Evolutionary Theory}, vol.~1,
  no.~1, pp.~1--30, 1973.

\bibitem{WEIT05}
J.~Weitz, H.~Hartman, and S.~Levin, ``{Coevolutionary arms races between
  bacteria and bacteriophage},'' {\em Proceedings of the National Academy of
  Sciences}, vol.~102, no.~27, pp.~9535--9540, 2005.

\bibitem{brockhurst2003population}
M.~Brockhurst, A.~Morgan, P.~Rainey, and A.~Buckling, ``{Population mixing
  accelerates coevolution},'' {\em Ecology Letters}, vol.~6, no.~11,
  pp.~975--979, 2003.

\bibitem{brockhurst2004effect}
M.~Brockhurst, P.~Rainey, and A.~Buckling, ``{The effect of spatial
  heterogeneity and parasites on the evolution of host diversity.},'' {\em
  Proceedings of the Royal Society B: Biological Sciences}, vol.~271, no.~1534,
  p.~107, 2004.

\bibitem{fussmann2007eco}
G.~Fussmann, M.~Loreau, and P.~Abrams, ``{Eco-evolutionary dynamics of
  communities and ecosystems},'' {\em Ecology}, vol.~21, pp.~465--477, 2007.

\bibitem{gandon2008host}
S.~Gandon, A.~Buckling, E.~Decaestecker, and T.~Day, ``{Host-parasite
  coevolution and patterns of adaptation across time and space},'' {\em Journal
  of Evolutionary Biology}, vol.~21, no.~6, pp.~1861--1866, 2008.

\bibitem{brockhurst2010using}
M.~Brockhurst, ``{Using Microbial Microcosms to Study Host--parasite
  Coevolution},'' {\em Evolution: Education and Outreach}, pp.~1--5, 2010.

\bibitem{paterson2010antagonistic}
S.~Paterson, T.~Vogwill, A.~Buckling, R.~Benmayor, A.~Spiers, N.~Thomson,
  M.~Quail, F.~Smith, D.~Walker, B.~Libberton, {\em et~al.}, ``{Antagonistic
  coevolution accelerates molecular evolution},'' {\em Nature}, 2010.

\bibitem{hillesland2010rapid}
K.~Hillesland and D.~Stahl, ``{Rapid evolution of stability and productivity at
  the origin of a microbial mutualism},'' {\em Proceedings of the National
  Academy of Sciences}, vol.~107, no.~5, p.~2124, 2010.

\bibitem{KOND03}
M.~Kondoh, ``{Foraging adaptation and the relationship between food-web
  complexity and stability},'' 2003.

\bibitem{ACKL04}
G.~Ackland and I.~Gallagher, ``{Stabilization of large generalized
  Lotka-Volterra foodwebs by evolutionary feedback},'' {\em Physical review
  letters}, vol.~93, no.~15, p.~158701, 2004.

\bibitem{WANG10}
Z.~Wang and N.~Goldenfeld, ``Stochastic model of antagonistic coevolution in a
  predator-prey model.'' unpublished.

\bibitem{AXEL81}
R.~Axelrod and W.~Hamilton, ``{The evolution of cooperation.},'' {\em Science},
  vol.~211, no.~4489, pp.~1390--1396, 1981.

\bibitem{SMIT82}
J.~Smith, {\em {Evolution and the Theory of Games}}.
\newblock Cambridge Univ Pr, 1982.

\bibitem{NOWA04}
M.~A. Nowak and K.~Sigmund, ``{Evolutionary Dynamics of Biological Games},''
  {\em Science}, vol.~303, no.~5659, pp.~793--799, 2004.

\bibitem{NOWA06}
A.~Nowak~Martin, {\em {Evolutionary Dynamics: Exploring the Equations of
  Life}}.
\newblock Cambridge, 2006.

\bibitem{MCGI07}
B.~McGill and J.~Brown, ``{Evolutionary Game Theory and Adaptive Dynamics of
  Continuous Traits},'' {\em Annu. Rev. Ecol. Evol. Syst}, vol.~38,
  pp.~403--35, 2007.

\bibitem{TURN99}
P.~Turner and L.~Chao, ``{Prisoner's dilemma in an RNA virus.},'' {\em Nature},
  vol.~398, no.~6726, pp.~441--443, 1999.

\bibitem{BROW01}
S.~Brown, ``{Collective action in an RNA virus},'' {\em Journal of Evolutionary
  Biology}, vol.~14, no.~5, pp.~821--828, 2001.

\bibitem{KIRK04}
B.~Kirkup and M.~Riley, ``{Antibiotic-mediated antagonism leads to a bacterial
  game of rock--paper--scissors in vivo},'' {\em Nature}, vol.~428, no.~6981,
  pp.~412--414, 2004.

\bibitem{GORE09}
J.~Gore, H.~Youk, and A.~Van~Oudenaarden, ``{Snowdrift game dynamics and
  facultative cheating in yeast},'' {\em Nature}, vol.~459, no.~7244,
  pp.~253--256, 2009.

\bibitem{NASH50}
J.~Nash, ``{Equilibrium points in n-person games},'' {\em Proceedings of the
  National Academy of Sciences of the United States of America}, pp.~48--49,
  1950.

\bibitem{AKIY2000}
E.~Akiyama and K.~Kaneko, ``{Dynamical systems game theory and dynamics of
  games},'' {\em Physica D: Nonlinear Phenomena}, vol.~147, no.~3-4,
  pp.~221--258, 2000.

\bibitem{AKIY02}
E.~Akiyama and K.~Kaneko, ``{Dynamical systems game theory II:: A new approach
  to the problem of the social dilemma},'' {\em Physica D: Nonlinear
  Phenomena}, vol.~167, no.~1-2, pp.~36--71, 2002.

\bibitem{SATO02}
Y.~Sato, E.~Akiyama, and J.~Doyne~Farmer, ``{Chaos in learning a simple
  two-person game},'' in {\em Proceedings of the National Academy of Science},
  vol.~99, pp.~4748--4751, 2002.

\bibitem{ZIMM01}
M.~Zimmermann, V.~Egu{\i}luz, and M.~Miguel, ``{Cooperation, adaptation and the
  emergence of leadership Economics with Heterogeneous Interacting Agents vol
  503},'' {\em Lecture Notes in Economics and Mathematical Systems},
  pp.~73--86, 2001.

\bibitem{EBEL02}
H.~Ebel and S.~Bornholdt, ``Coevolutionary games on networks,'' {\em Phys. Rev.
  E}, vol.~66, no.~5, pp.~056118/1--8, 2002.

\bibitem{PERC09}
M.~Perc and A.~Szolnoki, ``{Coevolutionary games--A mini review},'' {\em
  BioSystems}, vol.~99, pp.~109--125, 2009.

\bibitem{SATO05}
Y.~Sato, E.~Akiyama, and J.~Crutchfield, ``{Stability and diversity in
  collective adaptation},'' {\em Physica D: Nonlinear Phenomena}, vol.~210,
  no.~1-2, pp.~21--57, 2005.

\bibitem{FARM99}
J.~Farmer and A.~Lo, ``{Frontiers of finance: Evolution and efficient
  markets},'' {\em Proceedings of the National Academy of Sciences of the
  United States of America}, vol.~96, no.~18, pp.~9991--9992, 1999.

\bibitem{HENS05}
T.~Hens and K.~Schenk-Hopp{\'e}, ``{Evolutionary finance: introduction to the
  special issue},'' {\em Journal of mathematical economics}, vol.~41, no.~1-2,
  pp.~1--5, 2005.

\bibitem{MAY08}
R.~May, S.~Levin, and G.~Sugihara, ``{Ecology for bankers},'' {\em Nature},
  vol.~451, no.~21, pp.~893--895, 2008.

\bibitem{LURI43}
S.~Luria and M.~Delbr{\"u}ck, ``{Mutations of bacteria from virus sensitivity
  to virus resistance},'' {\em Genetics}, vol.~28, no.~6, pp.~491--511, 1943.

\bibitem{SNIE95}
P.~Sniegowski and R.~Lenski, ``{Mutation and adaptation: the directed mutation
  controversy in evolutionary perspective},'' {\em Annual Review of Ecology and
  Systematics}, vol.~26, no.~1, pp.~553--578, 1995.

\bibitem{GALH07}
R.~Galhardo, P.~Hastings, and S.~Rosenberg, ``{Mutation as a stress response
  and the regulation of evolvability},'' {\em Critical Reviews in Biochemistry
  and Molecular Biology}, vol.~42, no.~5, pp.~399--435, 2007.

\bibitem{dunny2007peptide}
G.~Dunny, ``{The peptide pheromone-inducible conjugation system of Enterococcus
  faecalis plasmid pCF10: cell--cell signalling, gene transfer, complexity and
  evolution},'' {\em Philosophical Transactions of the Royal Society B:
  Biological Sciences}, vol.~362, no.~1483, pp.~1185--1193, 2007.

\bibitem{MANS10}
J.~Manson, L.~Hancock, and M.~Gilmore, ``{Mechanism of chromosomal transfer of
  Enterococcus faecalis pathogenicity island, capsule, antimicrobial
  resistance, and other traits},'' {\em Proceedings of the National Academy of
  Sciences}, vol.~107, no.~27, pp.~12269--12274, 2010.

\bibitem{SCHRO54}
E.~Schr\"odinger, {\em Nature and the Greeks}.
\newblock Cambridge University Press, 1954.

\bibitem{FRAU99}
H.~Frauenfelder, P.~Wolynes, and R.~Austin, ``{Biological physics},'' {\em
  Reviews of Modern Physics}, vol.~71, no.~2, pp.~419--430, 1999.

\end{thebibliography}

\end{document}